\documentclass[twocolumn,preprintnumbers,amsmath,amssymb,balancelastpage]{revtex4}

\pdfoutput=1

\usepackage[utf8]{inputenc}
\usepackage[colorlinks=true,linkcolor=blue,urlcolor=blue,filecolor=black,citecolor=red,]{hyperref}

\usepackage{putexPRL}

\usepackage{enumerate}
\usepackage{mathtools}
\usepackage{graphicx}
\usepackage{placeins} 
\usepackage{float} 

\newcommand{\es}[2]{
    \begin{equation}\label{#1}
        \begin{split} #2 \end{split}
    \end{equation}
}

\newcommand{\Z}{\mathbb{Z}}

\newcommand{\cO}{\mathcal{O}}

\usepackage{tabularx}
\newcolumntype{Y}{>{\centering\arraybackslash}X}
\newcolumntype{C}[1]{>{\centering\arraybackslash}p{#1}}
\newenvironment{doublerule}{
    \noindent\hrule
    \vspace*{2pt}
    \noindent\hrule
    \vspace*{2pt}
}{
    \vspace*{2pt}
    \noindent\hrule
    \vspace*{2pt}
    \noindent\hrule
}

\definecolor{martenorange}{rgb}{0.91, 0.41, 0.17}
\newcommand{\red}{\color{red}}

\begin{document}

\title{Bootstrapping the Simplest Deconfined Quantum Critical Point}
\author{Shai M.~Chester}
\affiliation{
    Abdus Salam Centre for Theoretical Physics, Imperial College London, London SW7 2AZ, UK
}
\author{Alessandro Piazza}
\affiliation{
    SISSA, Via Bonomea 265, I-34136 Trieste, Italy \\
    INFN, Sezione di Trieste, Via Valerio 2, I-34127 Trieste, Italy
}
\author{Marten Reehorst}
\affiliation{
    Department of Mathematics, King’s College London, Strand, London, WC2R 2LS, UK
}
\author{Ning Su}
\affiliation{
    Department of Physics, Massachusetts Institute of Technology, Cambridge, MA 02139, USA \\
    Walter Burke Institute for Theoretical Physics, Caltech, Pasadena, California 91125, USA
}

\begin{abstract}
    We study the $N=3$ case of the $CP^{N-1}$ model, which is a field theory of $N$ complex scalars in \( 3d \) coupled to an Abelian gauge field with $SU(N)\times U(1)$ global symmetry.
    Recent evidence suggests the $N=2$ theory is not critical, which makes the $N=3$ theory the simplest possibility of deconfined quantum criticality.
    We apply the conformal bootstrap to correlators of charge $q=0,1,2$ scalar operators under the $U(1)$ symmetry, which gives us access also to $q=3,4$ operators.
    After imposing that only the lowest $q=0,1,2$ scalar operators are relevant, we find that the bootstrap bounds are saturated by the large $N$ prediction for $q=1,2,3,4$ scalar monopole operator scaling dimensions, which were shown earlier to be accurate even for small $N$, as well as a lattice prediction for the $q=0$ non-monopole scalar operator.
    We also predict the scaling dimensions of the lowest spinning monopole operators, which we match to the large charge prediction for spinning operators.
    This suggests that the critical $CP^{2}$ model is described by this bootstrap bound.
\end{abstract}

\maketitle
\nopagebreak

\section{Introduction}

Deconfined quantum critical points (DQCPs) describe the continuous phase transition between one phase that preserves a symmetry $H$ but breaks another symmetry $H'$, and a second phase that preserves $H'$ but breaks $H$~\cite{Senthil:2003eed,2004PhRvB..70n4407S}. The phase transition is described by a conformal field theory (CFT) with symmetry $H\times H'$. The simplest example of a DQCP is quantum electrodynamics in \( 2+1 \) dimensions (QED3) with $N>1$ matter fields (either fermions or bosons), where $H=U(1)$ and $H'=SU(N)$, and the gauge fields are emergent at the critical point (i.e.\ deconfined). At large $N$, it can be proven that this theory flows to a CFT in the IR, and observables can be computed in a $1/N$ expansion~\cite{PhysRevLett.32.292,Appelquist:1988sr}. However, condensed matter realizations of this theory exist only for small $N$~\cite{Senthil:2023vqd}, where the theory is strongly coupled, which makes it challenging to compute observables or even determine if the theory flows to a CFT.

The simplest putative DQCP would have $N=2$ bosons or fermions. The conformal bootstrap was used in~\cite{Li:2018lyb} to show that the $N=2$ fermionic theory is likely not conformal~\footnote{
    In particular, if the theory were conformal, then it would be expected to have an enhanced $O(4)$ symmetry~\cite{Karch:2016sxi,Seiberg:2016gmd,Wang:2017txt,Akhond:2019ued}, but the bootstrap puts bounds on the scaling dimension of the order parameter that exclude the lattice estimate~\cite{Qin:2017cqw} by a huge margin.
}, and instead the theory is believed to spontaneously break $SU(2)\times U(1)\to U(1)$~\cite{Chester:2024waw,Dumitrescu:2024jko}. The bootstrap similarly rules out the possibility that the $N=2$ bosonic theory is critical~\cite{Poland:2018epd,Li:2018lyb}, i.e.\ that it is a CFT with one relevant $SU(2)\times U(1)$ singlet~\footnote{
    In particular, the theory is believed to have an enhanced $SO(5)$ symmetry~\cite{Nahum:2015vka}, but the lattice estimate for the order parameter is ruled out by the bootstrap if one assumes there is just one relevant $SU(2)\times U(1)$ singlet, i.e.\ no relevant $SO(5)$ singlets.
}. Instead, the theory is believed to either be a tricritical CFT (i.e.\ two relevant singlets)~\cite{Chester:2023njo,Takahashi:2024xxd}, or be weakly first order~\cite{Zhou:2023qfi}. Since $N$ must be even in the parity preserving fermionic case~\footnote{For odd $N$, one must have non-zero Chern-Simons coupling.}, this leaves the $N=3$ bosonic theory as the simplest possible DQCP with one relevant singlet.

QED3 with $N$ bosons, also known as the $CP^{N-1}$ model~\footnote{
    This theory is also referred to as non-compact QED3 or \( NCCP^{N-1} \) model in condensed matter literature, which refers to the fact that the theory has an explicit $U(1)$ symmetry that forbids monopoles from being added to the action.
}, has been extensively studied with lattice simulations, but there is no agreement on the values of scaling dimensions, or even for what value of $N$ the theory flows to a CFT. A recent simulation of lattice Abelian-Higgs model suggested that the theory is conformal for $N=10$ but not for $N=4$~\cite{Bonati:2020jlm}, which suggests it is also not conformal for $N=3$~\footnote{
    The theory is believe to be conformal for all $N\geq N_\text{crit}$, so if $N_\text{crit}=4$, then $N=3$ would not be conformal.
}.
Several previous simulations of the Néel-VBS transition in quantum \( SU(N) \) anti-ferromagnets, using the JQ model, suggested that the theory is conformal for $N=3$ and higher values~\cite{2009PhRvB..80r0414L,2013PhRvB..88v0408H}, but gave different scaling dimensions as reviewed in Table~\ref{tab:scalar-monopoles-n3}.

\begin{table}[t]
    \renewcommand{\arraystretch}{1.2}
    \setlength{\tabcolsep}{5pt}
    \begin{doublerule}
        \begin{tabularx}{\columnwidth}{c | XXXXX }
          \( CP^{2} \) & \( \Delta_{1} \) & \( \Delta_{2} \) & \( \Delta_{3} \) & \( \Delta_{4} \) & \( \Delta_{0} \) \\ \hline
          Bootstrap & \( 0.755^{*} \) & \( 1.841(1) \) & \( 3.173(4) \) & \( 4.65(9) \) & \( 1.61(1) \) \\
          Large \( N \) & \( 0.755 \) & \( 1.81 \) & \( 3.10 \) & \( 4.59 \) & -- \\
          Lattice~\cite{2009PhRvB..80r0414L} & \( 0.71(4) \) & -- & -- & -- & \( 1.46(7) \) \\
          Lattice~\cite{2013PhRvB..88v0408H} & \( 0.785 \) & \( 2.0 \) & -- & -- & \( 1.28 \)
        \end{tabularx}
    \end{doublerule}
    \caption{Comparison of scaling dimensions $\Delta_q$ of the lowest dimension scalar operators with $U(1)$ charge $q$ in the \( CP^{2} \) model, as determined from the bootstrap study here, the large $N$ expansion for monopoles $q>0$, and lattice studies for both monopoles and the lowest non-monopole singlet $q=0$. The asterisk by $\Delta_1$ for bootstrap means we put it in to determine the others, while the bootstrap errors come from extrapolation in the bootstrap truncation parameter $\Lambda$.\label{tab:scalar-monopoles-n3}}
\end{table}

\begin{table}[t]
    \renewcommand{\arraystretch}{1.2}
    \setlength{\tabcolsep}{5pt}
    \begin{doublerule}
        \begin{tabularx}{\columnwidth}{c|XXXX}
          Bootstrap & $\ell=1$ & $\ell=2$  & $\ell=3$  & $\ell=4$ \\ \hline
          $q=1$     & $3.328(4)$      & $3.21(1)$                                    & $4.1(1)$     & $5.42(3)$                                    \\
          $q=2$     & $4.83(3)$       &$3.634(6)$        & $5.64(9)$    & $5.6(1)$          \\
          $q=3$     & $6.0(1)$        & $4.893(4)$                                   & $5.6(1)$    & $6.82(2)$          \\
          $q=4$     & --            & $6.1(2)$      & -- & $7.87(7)$ \\
          $q=0$     & $2$     & $3.76(2)$                                    & $4.328(2)$   & $5.0(1)$
        \end{tabularx}
    \end{doublerule}
    \caption{Scaling dimensions $\Delta_{q,\ell}$ of the lowest dimension operators with $U(1)$ charge $q\in\mathbb{Z}$ and spin $\ell>0$ as determined from the bootstrap study here, with errors estimates coming from the extrapolation in the bootstrap truncation parameter $\Lambda$. Note that odd spin $q=4$ operators exist, but do not appear in the correlators we bootstrap here. The \( \ell=1, q = 0 \) operator is the conserved $U(1)$ current which has exact dimension $\Delta=2$.\label{tab:spin-monopoles-n3}
    }
\end{table}

In this work, we will show evidence that the $CP^2$ model flows to a critical point, and predict the scaling dimensions of low-lying operators. We will make use of the fact that the large $N$ expansion of scaling dimensions $\Delta_q$ of the lowest dimension scalar operators with charge $q\in\mathbb{Z}$ under the $U(1)$~\cite{Metlitski:2008dw,Dyer:2015zha}, which are called monopole operators~\cite{Borokhov:2002cg}, has been shown to be extremely accurate for all values of $N$ by comparison to lattice predictions for small values of $N$~\cite{Dyer:2015zha,2013PhRvB..88v0408H,2012PhRvL.108m7201K}, and the duality to the critical $O(2)$ model for $N=1$~\cite{Chester:2022wur,Peskin:1977kp,PhysRevLett.47.1556}.

We then apply the conformal bootstrap to $U(1)$ invariant CFTs in \( 3d \) whose only relevant operators are scalars with $q=0,1,2$, as suggested for the $CP^2$ model from large $N$. This excludes the critical $O(2)$ model, which also has a relevant $q=3$ operator~\cite{Chester:2019ifh}. The bootstrap rigorously bounds the space of allowed scaling dimensions of these operators, as well as the $q=3,4$ operators that appear in the operator product expansion (OPE) of the $q=0,1,2$ operators. Physical theories often appear approximately~\footnote{
    We emphasize that physical theories do not exactly appear on the boundary of an allowed region formed by bootstrapping a finite amount of correlators, because this would then imply that adding further correlators to the bootstrap could not change the bound even slightly. Nonetheless, physical theories are very near the boundary, as the CFT data read off from the approximate solution to crossing at the boundary matches independent estimates to good accuracy in all the examples listed in the main text.
} on the boundary of allowed regions, as has been shown for many other models~\footnote{
    For instance, the critical $O(N)$ models~\cite{Kos:2013tga,Kos:2015mba,Sirois:2022vth}, $O(N)\times O(2)$ models~\cite{Reehorst:2024vyq}, $N=4$ fermionic QED3~\cite{Chester:2016wrc,Albayrak:2021xtd}, $N=2$ bosonic QED3~\cite{Chester:2023njo}, bosonic QED3 for very large $N$~\cite{He:2021xvg}, the 3-state Potts model~\cite{Rong:2017cow,Chester:2022hzt}, the Gross-Neveu-Yukawa model~\cite{Iliesiu:2017nrv,Erramilli:2022kgp}, and the $\mathcal{N}=1$ Ising model~\cite{Atanasov:2018kqw,Rong:2018okz,Atanasov:2022bpi}.
}. In our case, by looking at the point on the boundary given by the large $N$ value of $\Delta_1$ and then minimizing the scaling dimension $\Delta_0$ of the lowest singlet operator (which is not a monopole), we can read off the values of all the other scaling dimensions. As shown in Table~\ref{tab:scalar-monopoles-n3}, our results match the large $N$ estimates of the scalar monopole scaling dimensions $\Delta_{q}$ for $q=2,3,4$, as well as the lattice estimate in~\cite{2009PhRvB..80r0414L} for $\Delta_0$.

We also make predictions for the lowest spin $\ell$ monopole operators with scaling dimension $\Delta_{q,\ell}$ for ${\ell\leq 4}$, reported in Table~\ref{tab:spin-monopoles-n3}. We can compare these values to the large charge $q$ effective theory at small ${\ell\neq1}$~\cite{Hellerman:2015nra,Monin:2016jmo}, whose coefficients were fixed using the large $N$ expansion for $\ell=0$ in~\cite{DeLaFuente:2018uee}. The extrapolation is summarized in Table~\ref{tab:spin-monopoles-n3-largeq}. We find that these values match our bootstrap predictions for $\ell\leq q$, which is presumably where this small $\ell$ expansion is accurate. Curiously, the exact same pattern is observed for the critical $O(2)$ model, where bootstrap results from~\cite{Liu:2020tpf} match the large~$q$ effective theory from~\cite{Banerjee:2017fcx} for $\ell\leq q$, as summarized in Appendix~\ref{app:o2}~\footnote{
    The large charge expansion for the critical $O(2)$ model was previously compared against lattice predictions for $\ell=0$ in~\cite{Cuomo:2023mxg}.
}.

\begin{table}[t]
    \renewcommand{\arraystretch}{1.2}
    \setlength{\tabcolsep}{5pt}
    \begin{doublerule}
        \begin{tabularx}{\columnwidth}{c|XXXX}
          Large charge  &$\ell=0$ & $\ell=2$ & $\ell=3$ & $\ell=4$ \\ \hline
          $q=1$ & $0.750$  & ${\red 2.482}$ & ${\red 3.199}$ & ${\red3.912}$  \\
          $q=2$ & $1.803$  & $3.536$ & ${\red4.253}$ & ${\red4.966}$  \\
          $q=3$ & $3.093$  & $4.825$ & $5.543$ & ${\red6.256}$  \\
          $q=4$  & $4.583$ & $6.315$ & \( 7.032 \) & $7.745$ \\
        \end{tabularx}
    \end{doublerule}
    \caption{Scaling dimensions $\Delta_{q,\ell}$ of the lowest dimension scalar operators with $U(1)$ charge \( q \in \mathbb{Z} \) and spin $\ell$ as computed from the large $q$ effective theory for general $\ell\neq1$, with coefficients fixed by large $N$ for $\ell=0$. The red values for $\ell>q$ are where this small $\ell$ expansion is observed to be less accurate.\label{tab:spin-monopoles-n3-largeq}  
    }
\end{table}

\section{The $CP^{N-1}$ model}\label{CPN}

The Lagrangian of $N$ complex scalar fields $\phi_i$ coupled to an Abelian gauge field $A_\mu$ in \( 3d \) is
\begin{equation}\label{CPNAction}
    \resizebox{\linewidth}{!}{\(\displaystyle
        \mathcal{L}=\sum_{i=1}^N\left[|(\partial_\mu-i A_\mu)\phi^i|^2+m^2|\phi^i|^2\right]+u\Big[ \sum_{i=1}^N|\phi^i|^2\Big]^2 +\frac{F^2}{4e^2}  \, ,
        \)
    }
    \vspace{1em}
\end{equation}
where $F_{\mu\nu}\equiv \partial_\mu A_\nu-\partial_\nu A_\mu$ is the field strength. At large $N$, we can tune $m^2=0$ to get a critical theory in the IR~\footnote{
    One can also tune both $m^2$ and $u$ to zero to get a tricritical theory.
}. One can then decouple the quartic term by introducing a Hubbard-Stratonovich field $\sigma$ and replacing $ u \left( \sum_{i=1}^N |\phi^i|^2 \right)^2$ by $\sigma \sum_{i=1}^N |\phi^i|^2 - \sigma^2/4u$, such that the functional integral over $\sigma$ reproduces~\eqref{CPNAction}. Since the scaling dimension of $\sigma$ is $\Delta_0=2$ at large $N$, in the IR we set $e,u\to\infty$ to get the conformally invariant action
\es{CPNActionI}{
    \mathcal{L}=&\sum_{i=1}^{N}\left[|(\partial_\mu-i A_\mu)\phi^i|^2+\sigma|\phi^i|^2\right] \, .
}
This is in the same universality class as the $CP^{N-1}$ model, since shifting $\sigma$ by one gives a non-linear sigma model with $CP^{N-1}$ target space~\cite{coleman_1985}. The theory has a $SU(N)$ flavor symmetry that rotates the $\phi_i$, as well as a $U(1)$ topological symmetry whose current  $\epsilon_{\mu\nu\rho}F^{\nu\rho}$ is conserved due to the Bianchi identity. Including discrete groups, the faithful global symmetry for \( N > 2 \) is $(SU(N)/\mathbb{Z}_N\times U(1))\rtimes \mathbb{Z}_2^c$, where $ \mathbb{Z}_2^c$ is charge conjugation and \( \mathbb{Z}_N \) is the center of \( SU(N) \).

We can construct local operators out of $\phi_i$, $A_\mu$, and $\sigma$ that transform under $SU(N)$ but not $U(1)$. For instance the unique relevant singlet scalar operator is $\sigma$, whose scaling dimension is $\Delta_0=2$ at $N\to\infty$. The scaling dimension of these operators can be computed at large $N$ using Feynman diagrams to get $\Delta\sim N^0$ and the sub-leading correction~\cite{PhysRevLett.32.292,Kaul_2008,Benvenuti:2018cwd,DIVECCHIA1981719,PhysRevB.54.11953,Vasilev:1983uw}, but this expansion is not very accurate for small $N$~\footnote{
    The scaling dimensions were also computed using the $4-\epsilon$ expansion~\cite{PhysRevLett.32.292,Folk1996OnTC,PhysRevB.100.134507}, which is also not accurate for $\epsilon=1$.
}.

\subsection{Monopole operators}

We define monopole operators as local gauge invariant operators that transform under the $U(1)$ with charge ${q\in \mathbb{Z}}$. These operators are not built from fields in the action, but instead are defined as inserting magnetic flux $q=\frac{1}{2\pi}\int F$~\cite{Murthy:1989ps}~\footnote{
    We normalize $q$ as twice the value given in~\cite{Chester:2023njo,Dyer:2015zha}.
}. The lowest dimension monopoles are scalars and singlets under $SU(N)$.

Their scaling dimensions $\Delta_q$ are identified via the state-operator correspondence with the ground state energies on $S^2\times \mathbb{R}$ with $2\pi q$ magnetic flux through the $S^2$, which can be computed at large $N$ using a saddle point expansion~\cite{Borokhov:2002cg,Metlitski:2008dw} to get $\Delta_q\sim N$. This calculation was carried out to sub-leading order in~\cite{Dyer:2015zha}, and the results were found to be extremely accurate even at small $N$. We list some values extrapolated to $N=3$ in Table~\ref{tab:scalar-monopoles-n3}.

\subsection{Large $q$ effective theory}
\label{secLargeQ}

As shown in~\cite{Hellerman:2015nra,Monin:2016jmo}, operators in any CFT with large charge $q$ under a $U(1)$ global symmetry are described by an effective theory. This effective theory can be used to compute the scaling dimension $\Delta_{q,\ell}$ of the lowest dimension primary operator with charge $q$ and spin $\ell\ll q^{\frac{1}{2}}$ as
\es{largeq}{
    \Delta_{q,\ell}\hspace{-.03in}=\hspace{-.03in}c_{\frac32}q^{\frac32}+c_{\frac12} q^{\frac12}-0.0937\hspace{-.04in}+\hspace{-.02in}\sqrt{\frac{\ell(\ell+1)}{2}}\hspace{-.04in}+O(q^{-\frac12}),
}
except for $\ell=1$, where the formula describes a conformal descendant~\footnote{This can be seen from the fact that $\sqrt{\ell(\ell+1)/2}=1$ for $\ell=1$.}.
Note that the $q^0$ terms are universal for any theory and any $\ell\ll q^{\frac{1}{2}}$. For the $CP^{N-1}$ model, the lowest dimension monopoles are described by this effective theory. The large $N$ results from~\cite{Dyer:2015zha}, valid for $\ell=0$ and finite $q$, were used in~\cite{DeLaFuente:2018uee} to compute the theory-dependent coefficients $c_{\frac{3}{2}}$ and $c_{\frac{1}{2}}$ to sub-leading order in $N$ (and verify the universal \( q^{0} \) term), which for $N=3$ give
\es{c0c1}{
    c_{\frac{3}{2}}=0.4983\,, \qquad c_{\frac12}=0.3449\,.
}
We show the values of $\Delta_{q,\ell}$ from this expansion in Table~\ref{tab:spin-monopoles-n3-largeq}, which for $\ell=0$ match the finite $q$ large $N$ results in Table~\ref{tab:scalar-monopoles-n3} to good accuracy.

\section{Conformal bootstrap}\label{num}

We will apply the conformal bootstrap to just the $O(2)\cong U(1)\rtimes \mathbb{Z}_2^c$ sector. We follow the setup for $O(2)$ invariant systems in~\cite{Chester:2019ifh}, which we review in more detail in Appendix~\ref{app:boot}. Four point functions of local scalar primary operators $\varphi_q(x)$ with charge \( q \) under a $U(1)$ global symmetry can be expanded in conformal blocks $g^{\Delta^-_{12},\Delta^-_{34}}_{\Delta,\ell}(u,v)$ as
\es{4point}{
    \hspace{-.1in}&\left\langle  \varphi_{q_1}(x_1)  \varphi_{q_2}(x_2)   \varphi_{q_3}(x_3)   \varphi_{q_4}(x_4)  \right\rangle=  \frac{x^{\Delta^-_{12}}_{24} x^{\Delta^-_{34}}_{14}}{x^{\Delta^-_{12}}_{14} x_{13}^{\Delta^-_{34}} }      \\
    \hspace{-.1in}&\times\frac{1} {x_{12}^{\Delta_{12}^+}x_{34}^{\Delta_{34}^+}} \sum_{\cO}\lambda_{\varphi_{q_1}\varphi_{q_2} \cO}\lambda_{\varphi_{q_3}\varphi_{q_4} \cO}T^{\mathbf{r}}_{\mathbf{r}_1\mathbf{r}_2\mathbf{r}_3\mathbf{r}_4}g^{\Delta^-_{12},\Delta^-_{34}}_{\Delta,\ell}(u,v),
}
where $\Delta^\pm_{ij}\equiv\Delta_i\pm\Delta_j$, \( x_{ij} = x_{i} - x_{j} \), and we define the conformal cross ratios as $ u \equiv \frac{{x_{12}^2x_{34}^2}}{{x_{13}^2x_{24}^2}}, v \equiv \frac{{x_{14}^2x_{23}^2}}{{x_{13}^2x_{24}^2}}$. The primary operators $\cO$ that appear in both OPEs $\varphi_{q_{1}}\times\varphi_{q_{2}}$ and ${\varphi_{q_{3}}\times\varphi_{q_{4}}}$ have scaling dimension $\Delta$, spin $\ell$, and transform in an irreducible representation $\bf r$ of \( O(2) \) that appears in both the tensor products ${\bf r}_1\otimes{\bf r}_2$ and ${\bf r}_3\otimes{\bf r}_4$. The global symmetry tensor structures are denoted by \( T^{\mathbf{r}}_{\mathbf{r}_1\mathbf{r}_2\mathbf{r}_3\mathbf{r}_4} \), whose explicit expression can be found in~\cite{Chester:2019ifh}. Finally, \( \lambda_{\varphi_{i}\varphi_{j}\cO} \) are the OPE coefficients. In the following we denote by \( \lambda_{ijk} \equiv \lambda_{\varphi_{i}\varphi_{j}\varphi_{k}} \) the OPE coefficients involving the ``external'' operators, whenever they are exchanged in the respective OPE. The LHS of~\eqref{4point} is invariant under swapping $ \varphi_{q_1}(x_1)$ with $ \varphi_{q_3}(x_3)$, but the block expansion on the RHS changes giving the crossing equations
\es{crossing}{
    0=& \sum_{\cO}\lambda_{\varphi_{q_1}\varphi_{q_2} \cO}\lambda_{\varphi_{q_3}\varphi_{q_4} \cO}T^{\mathbf{r}}_{\mathbf{r}_1 \mathbf{r}_2\mathbf{r}_3\mathbf{r}_4}v^{\frac{\Delta_{23}^+}{2}}g^{\Delta^-_{12},\Delta^-_{34}}_{\Delta,\ell}(u,v)\\
    &-\sum_{\cO}\lambda_{\varphi_{q_3}\varphi_{q_2} \cO}\lambda_{\varphi_{q_1}\varphi_{q_4}\cO}T^{\mathbf{r}}_{\mathbf{r}_3\mathbf{r}_2\mathbf{r}_1\mathbf{r}_4}u^{\frac{\Delta_{12}^+}{2}}g^{\Delta^-_{32},\Delta^-_{14}}_{\Delta,\ell}(v,u)\,.
}
By projecting this equation into a basis for the tensor structures $T^{\mathbf{r}}_{\mathbf{r}_1\mathbf{r}_2\mathbf{r}_3\mathbf{r}_4}$, one can get a finite set of crossing equations, each a function of $u,v$. One can then truncate the infinite dimensional vector space of functions of $u,v$ to the space of their derivatives at the crossing symmetric point up to a given order $\Lambda$~\footnote{
    One must also truncate the set of spins to some maximal value, but in practice this does not effect the numerical results as long as the value is large enough. For more details, see e.g.~\cite{Chester:2019wfx}.
}. By imposing the reality of the OPE coefficients that follows from unitarity, one can then use this finite set of crossing equations to compute bounds on the allowed scaling dimensions and OPE coefficients that appear in~\eqref{4point}, and these bounds monotonically shrink as $\Lambda$ is increased~\cite{Rattazzi:2008pe}. We use the \texttt{Navigator} algorithm to efficiently scan over the allowed region~\cite{Reehorst:2021ykw}.

\begin{figure}
	\centering
	\includegraphics[width=\columnwidth]{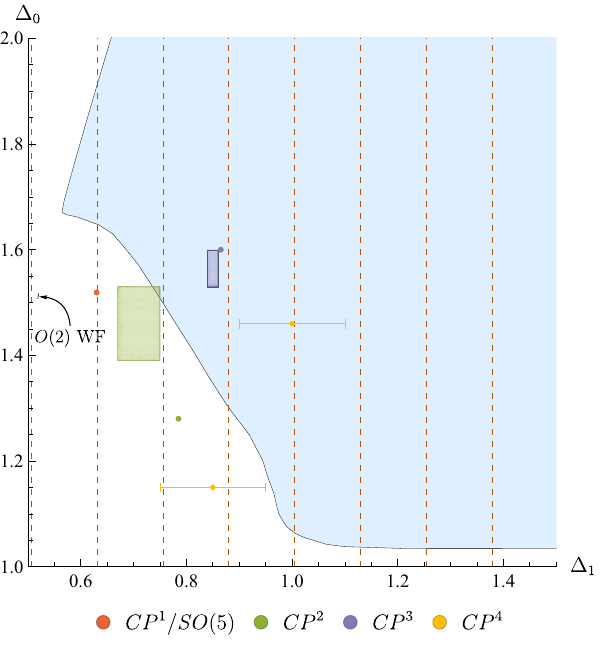}
	\caption{Allowed region for $(\Delta_1,\Delta_0)$ in \( 3d \) CFTs with $O(2)$ global symmetry and only one relevant $q=0,1$ operators, computed with $\Lambda=19$. These assumptions allow for the critical $O(2)$ Wilson-Fisher model, contained in the tiny island indicated by the arrow. They also allow for the $CP^{N-1}$ model for $N>2$. We show lattice results for $N=3,4,5$ from~\cite{2009PhRvB..80r0414L,2013PhRvB..88v0408H} and~\cite{2013PhRvL.111m7202B}, where errors bars are shown where available, and otherwise are shown by colored dots. The dashed vertical lines correspond to the value for \( \Delta_{1} \) extrapolated from large \( N \) estimates~\cite{Dyer:2015zha} down to \( N=1,2,\ldots,8 \) from left to right. The red point denotes the scaling dimensions of the \( SO(5) \) bootstrap~\cite{Chester:2023njo} which is believed to describe the tricritical \( N=2 \) theory, and is ruled out by our assumption of just one relevant $U(1)$ singlet.\label{fig:vs-continent}
	}
\end{figure}

We first consider correlators of $ \varphi_0$ and $\varphi_1$ for $\Lambda=19$, which yields 7 crossing equations as given in~\cite{Kos:2015mba}. We assume that these are the only relevant scalar operators with their $U(1)$ charges, and scan over the OPE coefficient ratio $\lambda_{000}/\lambda_{110}$ to further shrink the bound. The allowed region in the space of $(\Delta_0,\Delta_1)$ is shown in Figure~\ref{fig:vs-continent}, which generalizes Figure 2 of~\cite{Kos:2015mba} by extending to larger values of $\Delta_1$~\footnote{
    In~\cite{Kos:2015mba}, $\Delta_1$ and $\Delta_0$ are denoted as $\Delta_\phi$ and $\Delta_s$, respectively.
}. Note that the critical $O(2)$ Wilson-Fisher model and the $CP^{N-1}$ model for $N>2$ (if critical) should appear in the allowed region, since they both have one relevant $q=0,1$ operators~\footnote{
    Recall that the $CP^{1}$ theory is believed to either have two $q=0$ relevant operators or to have a weakly first order transition.
}. In particular, while the critical $O(2)$ theory appears in a small ``island'' as shown first in~\cite{Kos:2015mba}, the lattice and large $N$ estimates for the $CP^{2}$ theory appear on the lower boundary of the ``continent'', which motivates us to minimize $\Delta_0$ to find this theory. Note that as $\Delta_1$ gets bigger, the lower bound on $\Delta_0$ goes to one, so with these assumptions the $CP^{N-1}$ model for higher $N$ cannot lie on this bound, as we expect $\Delta_0\approx 2$ for large $N$. In Appendix~\ref{app:N4-5}, we discuss results for $N=4,5$, which show some agreement with the large $N$ expansion for scalar monopole scaling dimensions.

\begin{figure}
	\centering
	\includegraphics[width=\columnwidth]{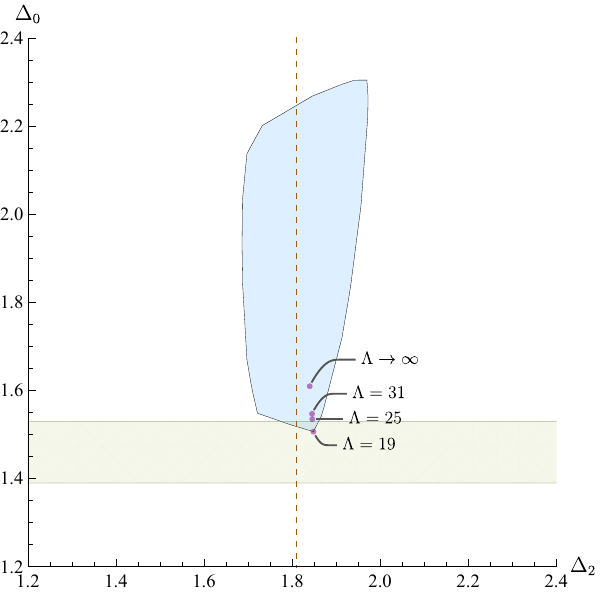}
	\caption{Allowed region for $(\Delta_2,\Delta_0)$ in \( 3d \) CFTs with $O(2)$ symmetry and only one relevant $q=0,1,2$ operators, with $\Delta_1=0.755$, and computed with $\Lambda=19$. The green band is the lattice result for \( \Delta_{0} \) from~\cite{2009PhRvB..80r0414L}, while the dashed vertical line is the large \( N \) value of \( \Delta_{2} \) extrapolated to \( N = 3 \). The purple dots are the result of \( \Delta_{0} \) minimization for some values of \( \Lambda \), together with the \( \Lambda \to \infty \) extrapolation.
	}
	\label{fig:vst-slice}
\end{figure}

We next consider correlators of $ \varphi_0$, $\varphi_1$, and $ \varphi_2$ for ${\Lambda=19}$, which yields 22 crossing equations as given in~\cite{Go:2019lke,Chester:2019ifh}. We now assume that these are the only relevant operators, by imposing that all $q=3,4$ scalar operators are irrelevant. This excludes the critical $O(2)$ model, which has a relevant $q=3$ operator~\cite{Chester:2019ifh}. The output of the bootstrap is an allowed region in the 6-dimensional space
$\big(\Delta_0,\Delta_ 1, \Delta_2 , \lambda_{000}/\lambda_{110}, \lambda_{220}/\lambda_{110}, \lambda_{112}/\lambda_{110}\big)$. We look at the point on the boundary of this region given by inputting the value $\Delta_1=0.755$, as given by large $N$, and minimizing $\Delta_0$ as suggested by the previous plot. In Figure~\ref{fig:vst-slice} we look at a cross section of this allowed region with $\Delta_1=0.755$ in the space of $(\Delta_0,\Delta_2)$, which shows that minimizing $\Delta_0$ corresponds to a corner of this allowed region.

We can then read off all CFT data that appears in the approximate solution to crossing at the boundary of this allowed region. We report the values of the lowest dimension $q=0,2,3,4$ scalar operators in Table~\ref{tab:scalar-monopoles-n3}, and compare them to large $N$ and lattice estimates. To obtain these values, we ran the bootstrap for $\Lambda=19$ to $\Lambda=31$ in steps of two, and then extrapolated to $\Lambda \to \infty$, so the error shown in our results comes from this extrapolation (see Appendix~\ref{app:N3} for more details). We also report some of the lowest dimension operators with higher spin in Table~\ref{tab:spin-monopoles-n3}. We do not report higher twist operators, since these are less accurate.

\section{Discussion}\label{conc}

In this work we bootstrapped the $CP^2$ model, which is the simplest DQCP. In particular, we showed that the point on the boundary of the allowed region of $U(1)$ CFTs in \( 3d \) given by inputting $\Delta_1$ as computed from the large $N$ expansion of the $CP^{N-1}$ model extrapolated to $N=3$, gives $\Delta_q$ for $q=2,3,4$ that match their large $N$ values, as well as $\Delta_0$ that matches the lattice estimate from~\cite{2009PhRvB..80r0414L}. We also made predictions for $\Delta_{q,\ell}$ for $\ell>0$, which match the large $q$ effective theory as computed using the large $N$ expansion in~\cite{DeLaFuente:2018uee} for $\ell\leq q$.

It is striking that the large $q$ expansion matches our bootstrap results for every $\ell\leq q$ that we computed, as strictly speaking this expansion should only be valid for $\ell\ll q^{\frac{1}{2}}$. A similar successful match was also observed for the critical $O(3)$ model in~\cite{Rong:2023owx} for $\ell=2$ and $q=1,\dots,6$. In Appendix~\ref{app:o2}, we compare the bootstrap results from~\cite{Liu:2020tpf} to the large $q$ expansion for the critical $O(2)$ from~\cite{Banerjee:2017fcx}, where in this case lattice simulations for $\ell=0$ data were used to fix $c_{\frac32}$ and $c_{\frac12}$ in~\eqref{largeq}. We find the exact same pattern as observed here for $CP^2$. This suggests the general large $q$ effective theory works for a larger regime of $\ell$ than one might expect~\footnote{
    Note that the large charge expansion for larger values of $\ell$ was worked out in~\cite{Cuomo:2022kio}, but these expansions do not match bootstrap data either for the critical $O(2)$ model or for our theory, which might be because these large $q$ expansions are known to fewer orders than the small $\ell$ regime expansion considered here.
}.

Looking ahead, we would like to access the $SU(3)$ sector of the $CP^2$ model, perhaps by performing a mixed correlator study between adjoints of $SU(3)$ as well as the monopoles considered in this work. Unfortunately, we found that the simplest bootstrap setup, which only assumes one relevant $q=0,1,2$ and one adjoint, leads to weak bounds in the $SU(3)$ sector, which are far from the lattice estimates of lowest dimension adjoint operator in~\cite{2009PhRvB..80r0414L}. This is because in this setup, we have found the mixing between the adjoint correlator and the correlators in this work to be very weak. Indeed, it was known from earlier work that the \( SU(N) \) adjoint bootstrap yields weak bounds for small $N$~\cite{He:2021xvg,Manenti:2021elk}. It would be interesting to find well motivated physical assumptions that could overcome this problem.

It would also be nice to confirm our predictions for higher spin monopoles for general $q$ by generalizing the large $N$ and finite $q$ calculation in~\cite{Metlitski:2008dw,Dyer:2015zha} to non-lowest monopoles, as all the lowest monopoles in the $CP^{N-1}$ theory have zero spin. Such a calculation could also show evidence for the putative $SO(5)$ symmetry enhancement for the $N=2$ theory~\cite{Nahum:2015vka,Chester:2023njo,Takahashi:2024xxd}.

We also expect that some of our predictions can be compared to lattice simulations in the near future. In particular, $\Delta_{3}$ can be compared to the $\omega$ component of lattice simulations on the honeycomb lattice, while $\Delta_{2,2}$ could be compared to the $\omega$ component on the square lattice~\cite{Metlitski:2017fmd}.

We would also like to generalize our study to larger $N$, even just in the $U(1)$ sector. Note that the only specification of $N$ into our $U(1)$ bootstrap was by inputting the value of $\Delta_1$ as given from the large $N$ expansion, so one could in principle bootstrap $CP^{N-1}$ for larger $N$ by inputting a larger value of $\Delta_1$. As noted above though, the lower bound on $\Delta_0$ goes to one as $\Delta_1$ increases, so the $CP^{N-1}$ model cannot lie on this lower bound in that regime. Preliminary investigation suggests that increasing the gap above $\Delta_0$ above three, which was what we assumed in this work, might address this problem, but we have no independent way of determining how large that gap should be.

Finally, it would be nice to bootstrap other \( 3d \) non-supersymmetric gauge theories~\footnote{
    Precise bootstrap islands have been found for \( 3d \) supersymmetric gauge theories such as ABJM theory in~\cite{Agmon:2019imm,Chester:2024bij}, but this required the additional input of supersymmetric localization constraints.
}, such as QED3 with non-zero Chern-Simons coupling, or QCD3. It would also be nice to upgrade the existing bounds for low $N$ theories in this work and~\cite{Chester:2016wrc,Albayrak:2021xtd,Chester:2023njo} into precise islands, so that the bootstrap can numerically solve these theories in the same sense that the critical $O(N)$ model was solved in~\cite{ElShowk:2012ht,Kos:2016ysd,Kos:2015mba,Chester:2019ifh,PhysRevD.104.105013}.

\begin{acknowledgments}
    We thank Max Metlitski, Ribhu Kaul, Anders Sandvik, Matt Block, Gabriel Cuomo, Junyu Liu, David Simmons-Duffin and Nicola Dondi for useful conversations, and Max Metlitski for reviewing the manuscript. SMC is supported by the Royal Society under the grant URF\textbackslash R1\textbackslash 221310 and the UK Engineering and Physical Sciences Research council grant number EP/Z000106/1, and thanks the conference on Workshop on Higher-$d$ Integrability in Favignana, Sicily for hospitality as this project was completed. The work of MR is supported by the UK Research and Innovation (UKRI) under the UK government’s Horizon Europe funding Guarantee (grant number EP/X042618/1). The computations presented here were conducted in the SISSA HPC cluster Ulysses and in the Resnick High Performance Computing Center, a facility supported by Resnick Sustainability Institute at the California Institute of Technology.
\end{acknowledgments}

\bibliographystyle{ssg}
\bibliography{SU3.bib}

\newpage
\onecolumngrid
\appendix

\section*{Supplemental Materials}
\section{Numerical bootstrap details}
\label{app:boot}

We remind the reader that representations of \( O(2) \cong U(1) \rtimes \mathbb{Z}_{2} \) can be induced from the 1 dimensional complex representations of \( U(1) \), labeled by \( q \in \mathbb{Z} \). The \( q = 0 \) representation induces the trivial representation \( \mathbf{0}^{+} \) (singlet) and the sign representation \( \mathbf{0}^{-} \) (odd under \( \Z_{2} \)), both of which are 1 dimensional. The \( q > 0 \) representation induces a 2 dimensional representation \( \mathbf{q} \), isomorphic to \( q \oplus (-q) \) as \( U(1) \) representation and where \( \mathbb{Z}_{2} \) acts by swapping the two subspaces.
Our bootstrap setup involves all 4-point functions of scalar primary operators in the \( \mathbf{0}^{+}, \mathbf{1}, \mathbf{2} \) representations, which we denote by \( \varphi_{0}, \varphi_{1}, \varphi_{2} \) respectively. Primary operators \( \mathcal{O}_{\mathbf{r},\Delta,\ell} \) exchanged in the conformal block decomposition~\eqref{4point} are in representations of \( O(2) \) determined by the tensor product decomposition of the representations of the possible pairs of external operators. These are
\begin{equation}
    \begin{aligned}
      \mathbf{0}^{+} \otimes \mathbf{0}^{+} &= \mathbf{0}^{+} \, , \qquad \qquad & \mathbf{1} \otimes \mathbf{1} &= \mathbf{0}^{+} \oplus \mathbf{0}^{-} \oplus \mathbf{2} \, , \qquad \qquad & \mathbf{2} \otimes \mathbf{2} &= \mathbf{0}^{+} \oplus \mathbf{0}^{-} \oplus \mathbf{4}  \, ,  \qquad \qquad & \\
      \mathbf{0}^{+} \otimes \mathbf{1} &= \mathbf{1} \, , \qquad \qquad &
                                                                      \mathbf{0}^{+} \otimes \mathbf{2} &= \mathbf{2} \, , \qquad \qquad &
                                                                                                                                      \mathbf{1} \otimes \mathbf{2} &= \mathbf{1} \oplus \mathbf{3} \, .
    \end{aligned}
\end{equation}
Moreover, permutation symmetry of 3-point function implies that in the \( \varphi_{0} \times \varphi_{0}\), \( \varphi_{1} \times \varphi_{1} \) and \( \varphi_{2}\times \varphi_{2} \) OPEs the primary operators \( \mathcal{O}_{\mathbf{0}^{+}, \Delta, \ell} \), \( \mathcal{O}_{\mathbf{2}, \Delta, \ell} \) and \( \mathcal{O}_{\mathbf{4}, \Delta, \ell} \) are exchanged with even \( \ell \), while  \( \mathcal{O}_{\mathbf{0}^{-}, \Delta, \ell} \) are exchanged with odd \( \ell \). Instead, in the \( \varphi_{0} \times \varphi_{1} \), \( \varphi_{0} \times \varphi_{2} \) and \( \varphi_{1} \times \varphi_{2} \) OPEs the primary operators \( \mathcal{O}_{\mathbf{1}, \Delta, \ell} \), \( \mathcal{O}_{\mathbf{2}, \Delta, \ell} \) and \( \mathcal{O}_{\mathbf{3}, \Delta, \ell} \) are exchanged with any \( \ell \). The ``external'' operators appear themselves in the OPEs considered: \( \varphi_{0} \) appears in \( \varphi_{0} \times \varphi_{0} \), \( \varphi_{1}\times \varphi_{1} \) and \( \varphi_{2} \times \varphi_{2} \) with OPE coefficients \( \lambda_{000} \), \( \lambda_{110} \) and \( \lambda_{220} \) respectively; \( \varphi_{1} \) appears in  \( \varphi_{0} \times \varphi_{1} \) with coefficient \( \lambda_{101} = \lambda_{110} \); \( \varphi_{2} \) appears in \( \varphi_{1} \times \varphi_{1} \) and \( \varphi_{0} \times \varphi_{2} \) with coefficients \( \lambda_{112} \) and \( \lambda_{202} = \lambda_{220} \) respectively.

When sitting at points that saturate bootstrap bounds, i.e.\ when we minimize \( \Delta_{0} \) at fixed \( \Delta_{1} \), it's possible to obtain approximate solutions to the set of crossing equations using the Extremal Functional Method (EFM)~\cite{Poland:2010wg,El-Showk:2012vjm,Simmons-Duffin:2016wlq}. The EFM gives a non-rigorous estimate of the low lying spectrum and OPE coefficients of operators exchanged in OPEs involved in the crossing equations. With our bootstrap setup we can thus access the following operators: \( \mathcal{O}_{\mathbf{0}^{+}, \Delta, \ell} \) and  \( \mathcal{O}_{\mathbf{4}, \Delta, \ell} \) with \( \ell \) even; \( \mathcal{O}_{\mathbf{0}^{-}, \Delta, \ell} \) with \( \ell \) odd; \( \mathcal{O}_{\mathbf{1}, \Delta, \ell} \), \( \mathcal{O}_{\mathbf{2}, \Delta, \ell} \) and \( \mathcal{O}_{\mathbf{3}, \Delta, \ell} \) with any \( \ell \). The estimates from EFM are subject to a dependence on the dimension of the space of derivative functionals used in the bootstrap bound (parametrized by \( \Lambda \)) and typically we can obtain a reliable estimate only for the lowest twist operators. See Appendix~\ref{app:N3} and~\ref{app:N4-5} for the \( \Lambda \)-dependence of our bootstrap estimates of scaling dimensions.

We used the software \texttt{autoboot}~\cite{Go:2019lke} to obtain the crossing equations~\eqref{crossing} for correlators involving scalar operators \( \varphi_{0}, \varphi_{1} \), used for Figure~\ref{fig:vs-continent}, and for  correlators involving \( \varphi_{0}, \varphi_{1}, \varphi_{2} \), used for Figure~\ref{fig:vst-slice} as well as for the \texttt{Navigator} computations.
For the normalization convention of the OPE coefficients we refer to section 2 of the~\cite{Go:2019lke}.

The numerical bootstrap computations of this work were carried out using the \texttt{simpleboot} package~\cite{simpleboot}.
The computation of the conformal blocks was done using \texttt{scalar\_blocks}~\cite{scalarblocks}, while the semi-definite optimization was done using \texttt{SDPB}~\cite{Landry:2019qug}.
The feasibility bounds for scaling dimensions in Figures~\ref{fig:vs-continent} and~\ref{fig:vst-slice} are obtained with a delaunay triangulation in scaling dimensions and by performing a scan over the OPE ratios of external operators using the cutting surface algorithm introduced in~\cite{Chester:2019ifh}.
The constrained minimization of \( \Delta_{0} \) at fixed \( \Delta_{1} \) in the 5-dimensional space \( (\Delta_{0}, \Delta_{2}, \lambda_{000}/\lambda_{110}, \lambda_{220}/\lambda_{110}, \lambda_{112}/\lambda_{110}) \), reported in Tables~\ref{tab-raw-n3},~\ref{tab-raw-n4} and ~\ref{tab-raw-n5}, was carried out with the \texttt{Navigator} method, using a modified BFGS algorithm introduced in~\cite{Reehorst:2021ykw}.
Both the cutting surface and the BFGS algorithm are implemented in \texttt{simpleboot}.

The parameters used in the numerical computation of the conformal blocks are reported in Table~\ref{tab:blocks-param} for the various values of the bootstrap parameter \( \Lambda \) we explored. There, \( S_{\Lambda} \) denotes the following sets of spins included in the bootstrap constraints:
\begin{equation}\label{eq:spin-sets}
    \begin{aligned}
      S_{19} &= \{0,\ldots,26\}\cup\{49,52\} \, , \\
      S_{23} &= \{0,\ldots,30\}\cup\{38,49,52\} \, , \\
      S_{25} &= \{0,\ldots,30\}\cup\{38,49,52\} \, , \\
      S_{27} &= \{0,\ldots,31\}\cup\{33,34,37,38,41,42,45,46,49,50\} \, , \\
      S_{29} &= \{0,\ldots,31\}\cup\{33,34,37,38,41,42,45,46,49,50,55,56\} \, , \\
      S_{31} &= \{0,\ldots,31\}\cup\{33,34,37,38,41,42,45,46,49,50,55,56,59,60\} \, .
    \end{aligned}
\end{equation}
The parameters relevant for the semi-definite optimization are reported in Table~\ref{tab:sdpb}, both for the feasibility bounds and the Navigator run. For the feasibility bounds, we assumed a bounding box on the OPE ratios for the cutting surface algorithm of \( \big(\lambda_{000}/\lambda_{110}, \lambda_{220}/\lambda_{110}, \lambda_{112}/\lambda_{110}\big) \in [-10,10] \times [-10,10] \times [-10, 10] \).

For the extraction of the \( \Delta_{3} \) and \( \Delta_{4} \) scaling dimensions from the extremal functional, as well as for spinning operators, we used the \texttt{spectrum.py} package~\cite{Komargodski:2016auf,spectrum}.

\begin{table}[H]
    \centering
    \renewcommand{\arraystretch}{1.08}
    \setlength{\tabcolsep}{5pt}
    \begin{minipage}[t]{0.47\textwidth}
        \centering
        \begin{doublerule}
            \begin{tabularx}{\textwidth}{Y|cccc}
              \( \Lambda \) & \texttt{spin-ranges} & \texttt{poles} & \texttt{order} & \texttt{precision} \\ \hline
              19 & \( S_{19} \) & 14 & 56 & 1024 \\
              21 & \( S_{21} \) & 18 & 72 & 1024 \\
              23 & \( S_{23} \) & 18 & 72 & 1024 \\
              25 & \( S_{25} \) & 18 & 72 & 1024 \\
              27 & \( S_{27} \) & 20 & 80 & 1024 \\
              29 & \( S_{29} \) & 22 & 88 & 1024 \\
              31 & \( S_{31} \) & 24 & 96 & 1024
            \end{tabularx}
        \end{doublerule}
        \caption{Parameters of \texttt{scalar\_blocks} used for the computation of the conformal blocks. See~\eqref{eq:spin-sets} for the definition of the spin ranges.\label{tab:blocks-param}}
    \end{minipage}
    \hspace{2em}
    \begin{minipage}[t]{0.47\linewidth}
        \centering
        \begin{doublerule}
            \begin{tabularx}{\textwidth}{c|YY}
              Parameter & Feasibility & Navigator \\ \hline
              \texttt{dualityGapThreshold} & \( 10^{-20} \) & \( 10^{-30} \) \\
              \texttt{primalErrorThreshold} & \( 10^{-60} \) & \( 10^{-30} \) \\
              \texttt{dualErrorThreshold} & \( 10^{-60} \) & \( 10^{-30} \) \\
              \texttt{initialMatrixScalePrimal} & \( 10^{20} \) & \( 10^{20} \) \\
              \texttt{initialMatrixScaleDual}  & \( 10^{20} \) & \( 10^{20} \) \\
              \texttt{maxComplementarity} & \( 10^{100} \)  & \( 10^{100} \) \\
              \texttt{maxIterations} & 1000 & 1000 \\
              \texttt{precision} & 1024 & 1024 \\
              \texttt{detectPrimalFeasibleJump} & \texttt{true} & \texttt{false} \\
              \texttt{detectDualFeasibleJump}  & \texttt{true} & \texttt{false} \\
              \texttt{findPrimalFeasible} & \texttt{false} & \texttt{false} \\
              \texttt{findDualFeasible} & \texttt{false} & \texttt{false}
            \end{tabularx}
        \end{doublerule}
        \caption{Parameters of \texttt{SDPB} used for the semi-definite optimization.\label{tab:sdpb}}
    \end{minipage}
\end{table}

\section{Extended results for the $CP^{2}$ model}\label{app:N3}
In this section we describe in more detail the results obtained for the $CP^2$ model, which corresponds to studying the crossing equations~\eqref{crossing} for \( q = 0,1,2 \) operators at $\Delta_1=0.755$, in accordance with the large $N$ prediction.
In Tables~\ref{tab:scalar-monopoles-n3} and~\ref{tab:spin-monopoles-n3} in the main text we showed the estimates for scaling dimension of the the lowest dimensional monopoles.

\begin{figure}[H]
	\centering
	\begin{tabularx}{\textwidth}{YYY}
      \includegraphics[width=0.31\textwidth]{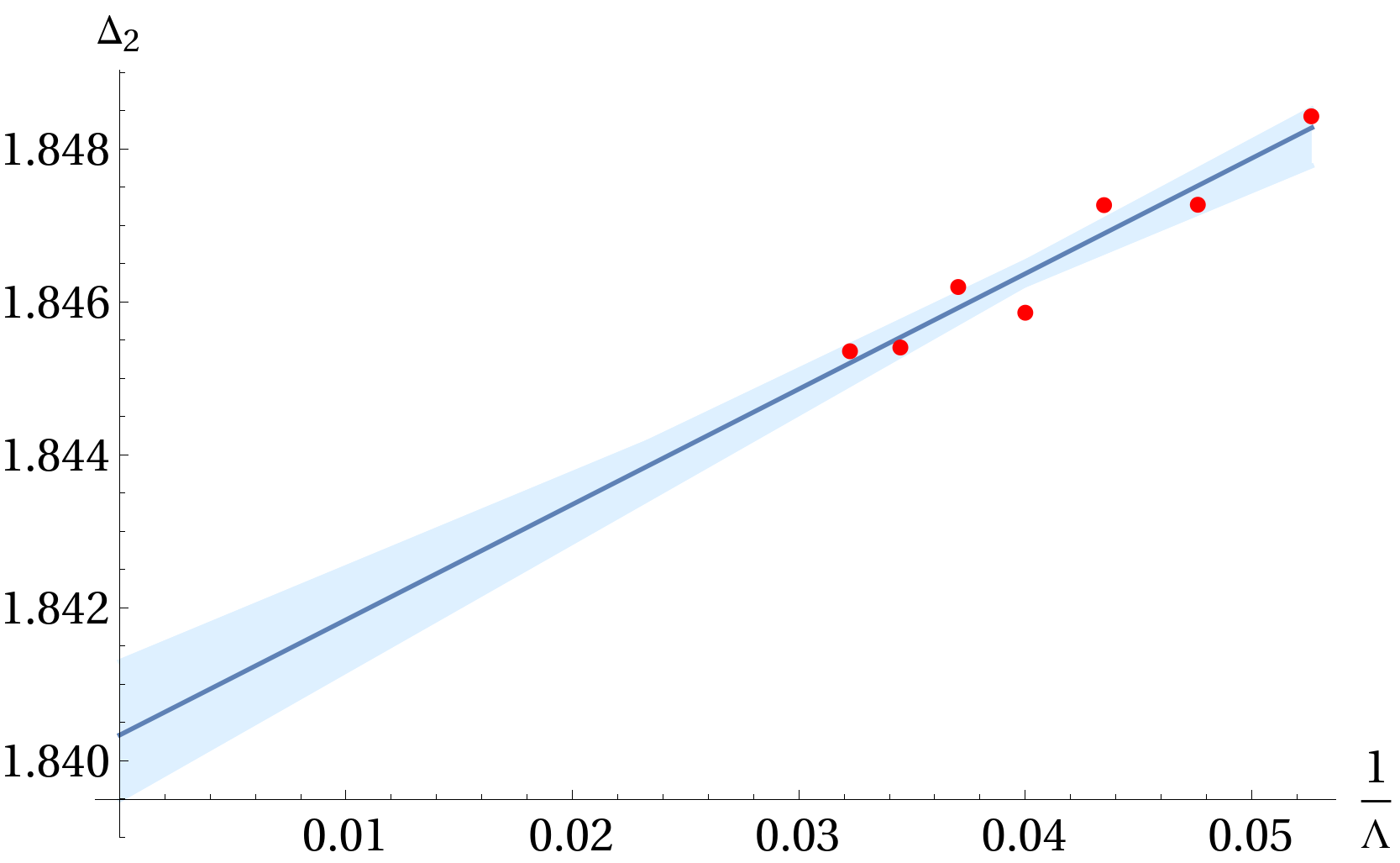}
      & \includegraphics[width=0.31\textwidth]{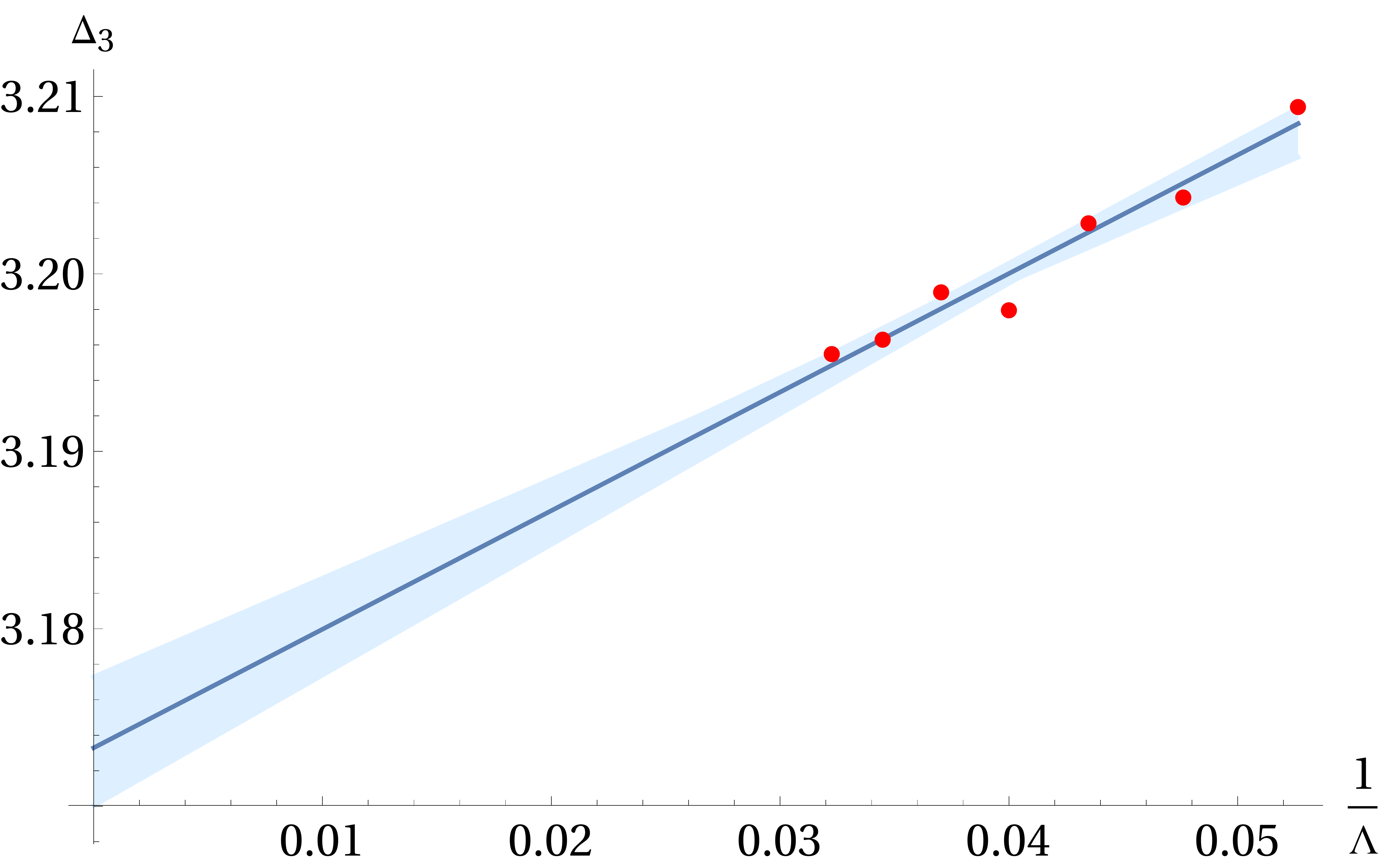}
      & \includegraphics[width=0.31\textwidth]{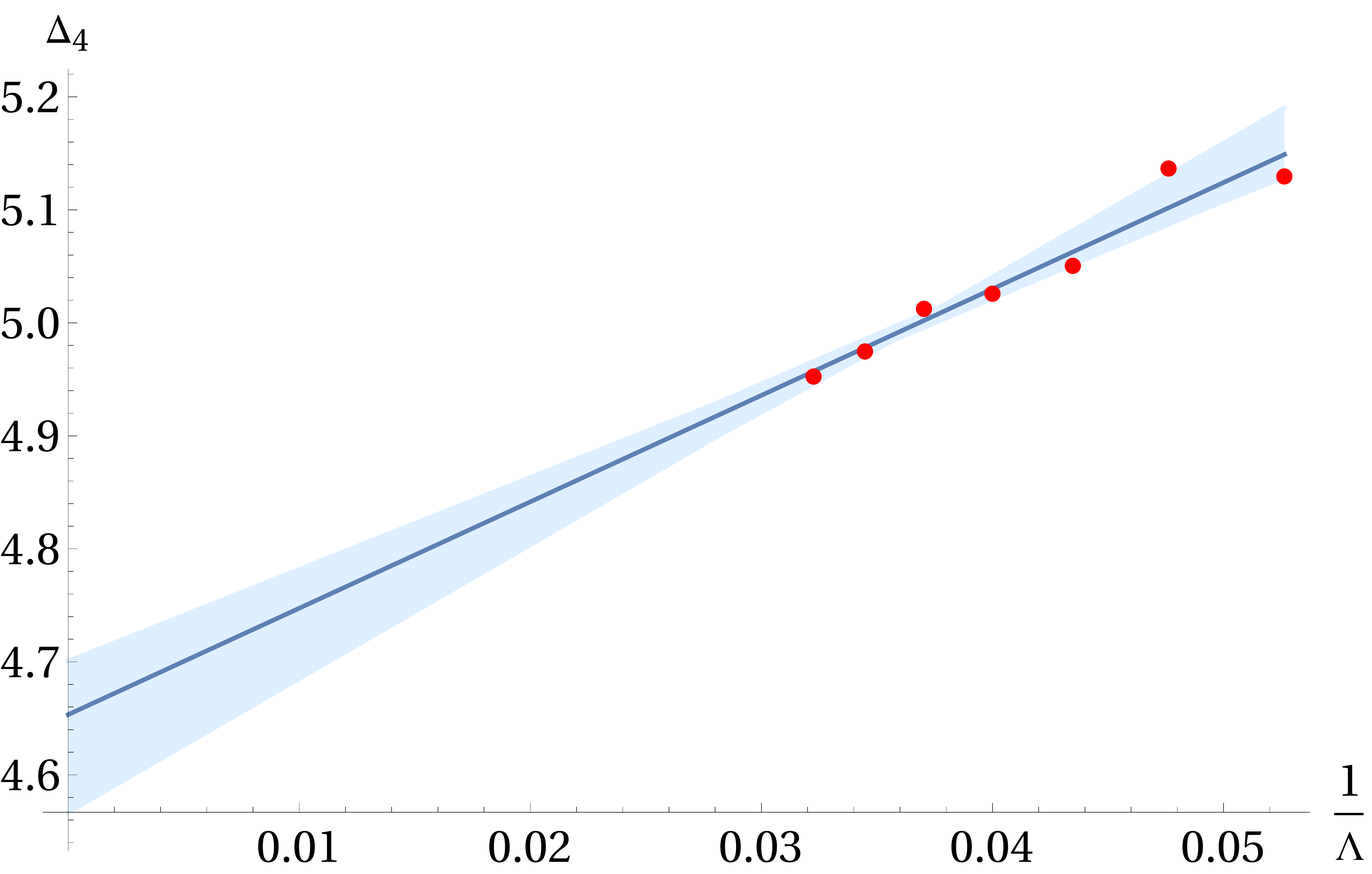}
	\end{tabularx}
	\caption{Linear least-squares fits in $1/\Lambda$ of the dimension of the lowest scalar monopoles in the $CP^{2}$ model. In the case of $\Delta_4$ we ignore a ``fake'' operator at the imposed gap of $3$. These fits are used to obtain the extrapolations at \( \Lambda \to \infty \) shown in Table~\ref{tab:scalar-monopoles-n3}. A blue band indicate the range of fits that can be obtained by omitting up to 2 points. The red dots indicate the dimensions (obtained by EFM except for $\Delta_2$ which is one of the variables in the search space).\label{fig:fitplots-n3}
	}
\end{figure}

These estimates are obtained as follows.
First we take the lowest dimension operator at multiple values of $\Lambda$, either as found using the EFM, or in the case of $\Delta_2$ and the ratio of OPE coefficients, the values found through a rigorous minimization.
Then we extrapolate our finite $\Lambda$ results to infinite $\Lambda$ assuming a linear fit in $1/\Lambda$.
Linear convergence in $1/\Lambda$ has been previously observed in a wide range of observables (for sufficiently large $\Lambda$).
Including a quadratic correction doesn't significantly improve the fit and does result in less physical behavior in the extrapolated regions.
To illustrate the fitting procedure we show the fits used in extrapolation for the scalar monopole dimension in Figure~\ref{fig:fitplots-n3}.
The data at finite values of $\Lambda$ used for the extrapolations is also shown in Table~\ref{tab-raw-n3}. The full data used for all the presented extrapolations can be found in the auxiliary file \texttt{QED3\_data.nb}. \\

It should be noted that no rigorous errors can be assigned to the values extracted by the EFM. Instead we include a non-rigorous estimate of the error in the extrapolated value in the form of the variation of the extrapolated value when leaving up to 2 points out (this method of estimating the predictive value of a prediction based on finite set of points is known as \emph{leave-p-out cross validation}, see for example Chapter 7.11 of~\cite{hastie2009elements}). \\

\begin{figure}[H]
	\centering
	\begin{tabularx}{\textwidth}{YYY}
      \includegraphics[width=0.31\textwidth]{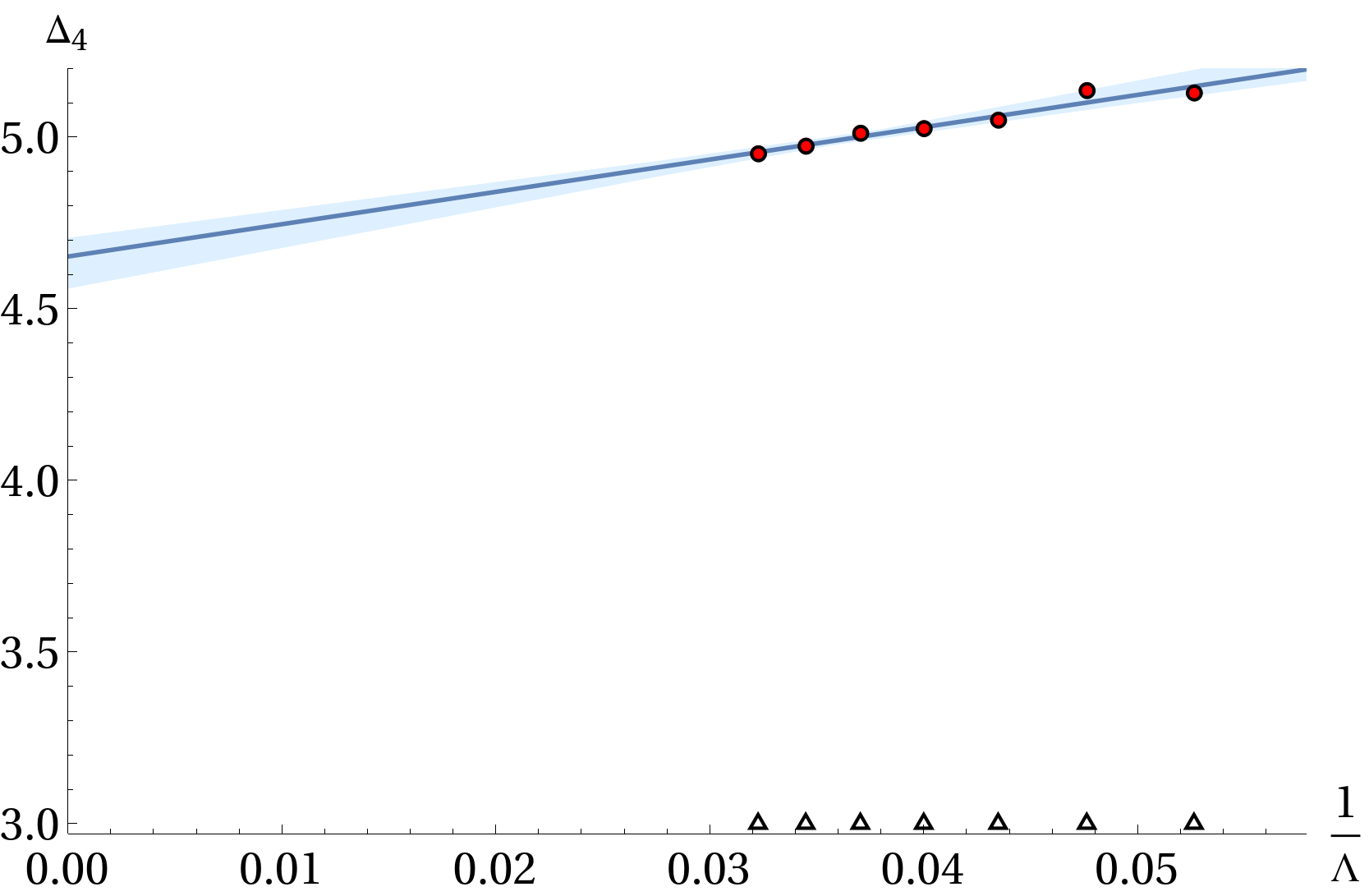}
      & \includegraphics[width=0.31\textwidth]{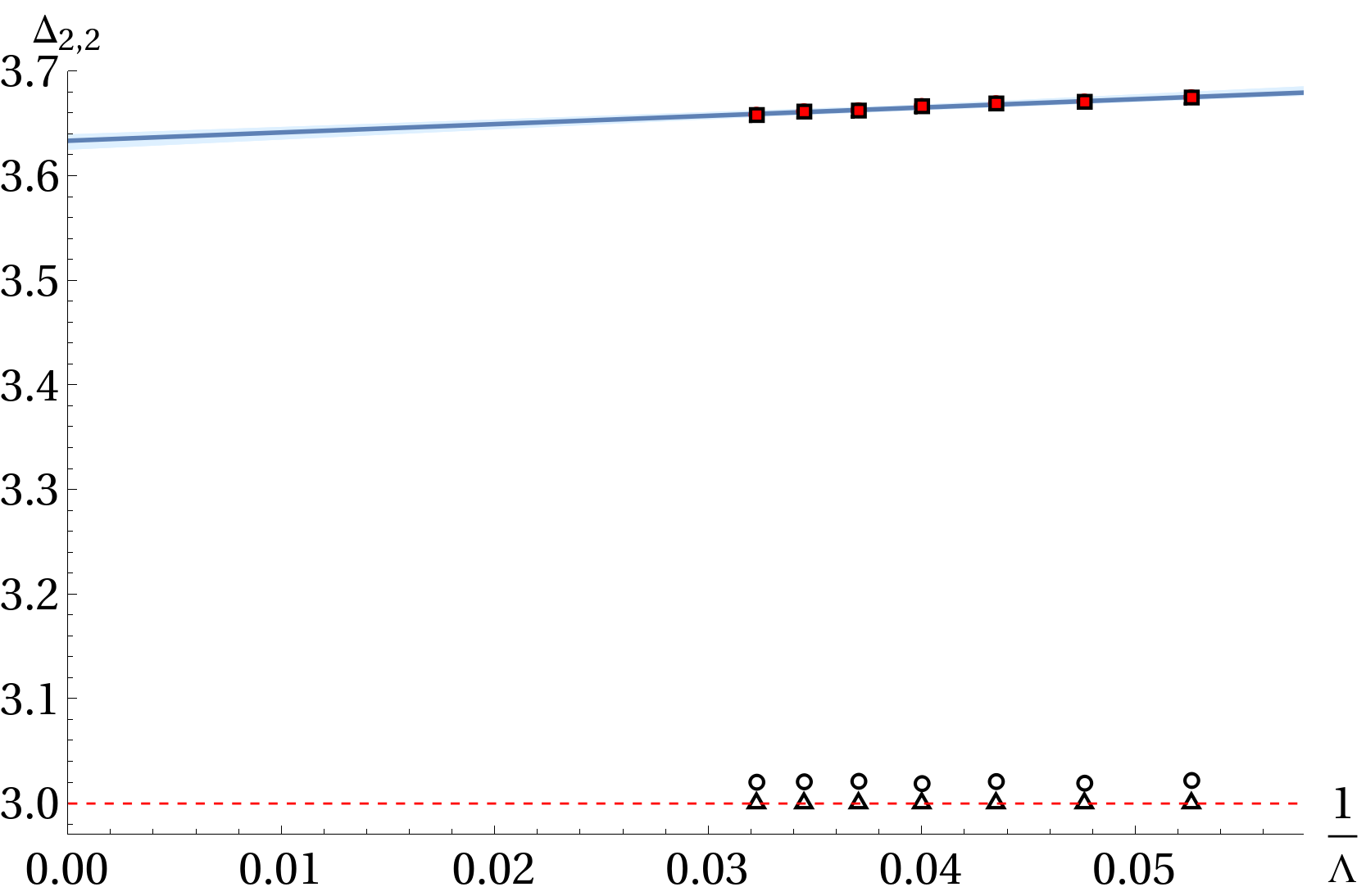}
      & \includegraphics[width=0.31\textwidth]{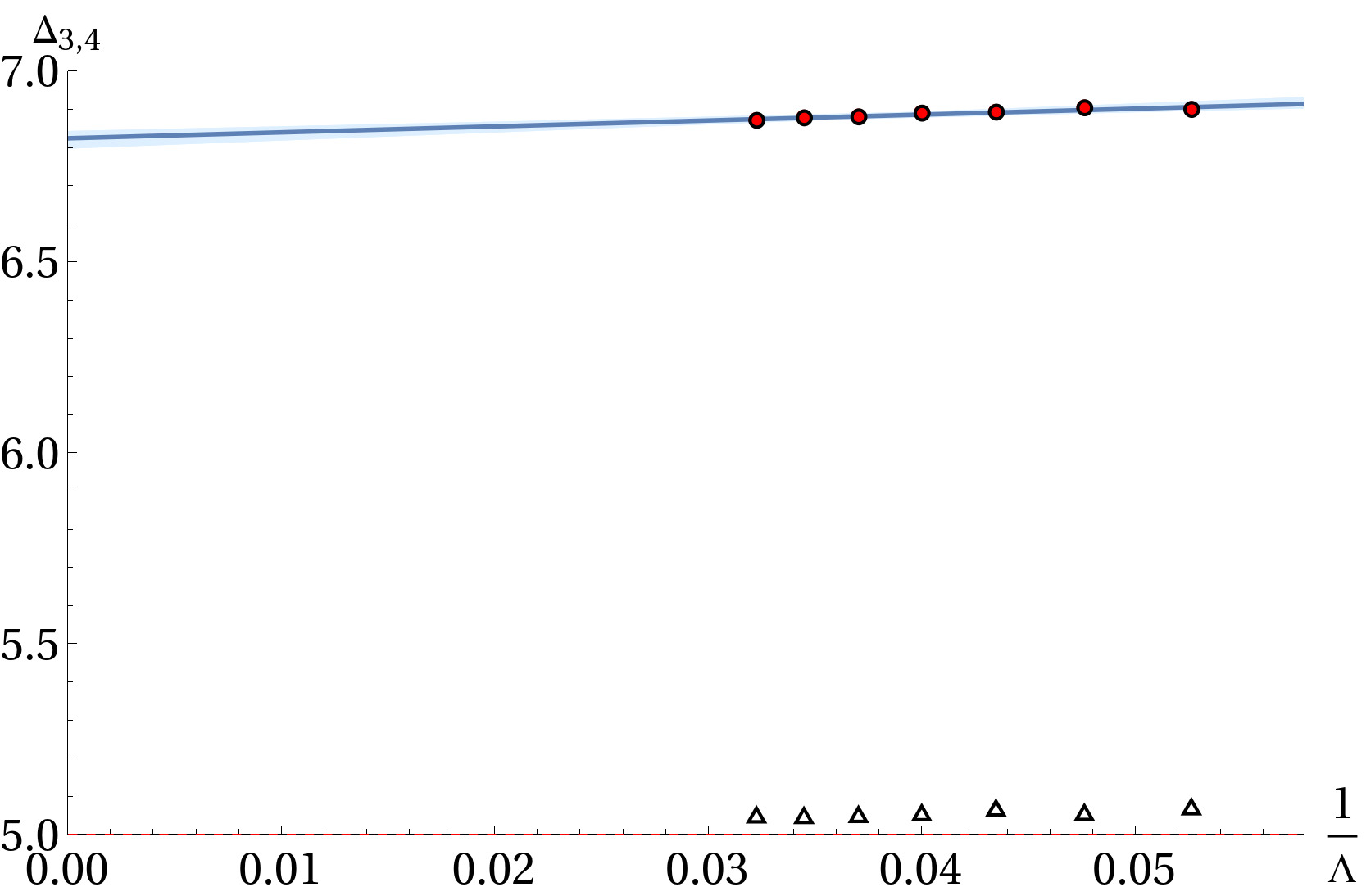}
	\end{tabularx}
	\caption{ Linear least-squares fits in $1/\Lambda$ of the suspected physical lowest dimensional operators in the $CP^{2}$ model. These fits are used to obtain the extrapolations at $\Lambda \to \infty$ shown in Tables~\ref{tab:scalar-monopoles-n3} and~\ref{tab:spin-monopoles-n3}. The operator dimensions obtained by EFM are plotted as black triangles (lowest dimensional operator), circles (next-lowest) and squares (next-to-next-to-lowest). The red dots indicate the dimensions used for extrapolations presented in Tables~\ref{tab:scalar-monopoles-n3} and~\ref{tab:spin-monopoles-n3}. Rather than using the lowest dimensional operators from the EFM as we did for the other sectors, in these case we treat operators at the imposed gaps or close to it as \emph{spurious}. This is supported by the fact that the OPE coefficients of these operators near the unitarity bound are significantly smaller than the OPE coefficients of the fitted operator and are getting smaller still as $\Lambda$ increases.\label{fig:fitplots-n3-special-cases}
	}
    \vspace{-2em}
\end{figure}

For some operators in Table~\ref{tab:spin-monopoles-n3} we interpret not the lowest but the second lowest operator detected in the EFM to be the true physical lowest operator. This is motivated because these spectra contain an extra operator very close to the unitary bound (and thus the assumed gap) with a significantly smaller OPE then the following operator. Moreover, this difference becomes larger as $\Lambda$ increases. This spurious operator could be related to the \textit{sharing effect}~\cite{Simmons-Duffin:2016wlq,Liu:2020tpf}. Such an operator close to the unitary bound is also not expected to exist according to the large-charge expansion. The choice of fit and the alternative points that we could have chosen to fit are illustrated in Figure~\ref{fig:fitplots-n3-special-cases}.

\begin{table}[H]
    \centering
    \setlength{\tabcolsep}{5pt}
    \begin{doublerule}
        \begin{tabularx}{\textwidth}{Y|@{\hskip 10 pt} XXXXXXXX}
          $CP^{2}$
          & $\Lambda=19$ & $\Lambda=21$ & $\Lambda=23$ & $\Lambda=25$
          & $\Lambda=27$ & $\Lambda=29$ & $\Lambda=31$ & $\Lambda \to \infty$ \\
          \hline
          $\Delta_0$
          & 1.507 & 1.519 & 1.528 & 1.535
          & 1.540 & 1.544 & 1.547 & 1.608 \\
          $\Delta_2$
          & 1.848 & 1.847 & 1.847 & 1.846
          & 1.846 & 1.845 & 1.845 & 1.841 \\
          $\Delta_3$
          & 3.209 & 3.204 & 3.203 & 3.198
          & 3.199 & 3.196 & 3.195 & 3.173 \\
          $\Delta_4$
          & 5.130 & 5.137 & 5.050 & 5.026
          & 5.012 & 4.975 & 4.952 & 4.653 \\
          $\lambda_{000}/\lambda_{110}$
          & 0.182 & 0.200 & 0.213 & 0.225
          & 0.234 & 0.241 & 0.246 & 0.344 \\
          $\lambda_{220}/\lambda_{110}$
          & 1.766 & 1.775 & 1.782 & 1.786
          & 1.789 & 1.792 & 1.794 & 1.836 \\
          $\lambda_{112}/\lambda_{110}$
          & 1.695 & 1.698 & 1.700 & 1.701
          & 1.703 & 1.703 & 1.704 & 1.717 \\
        \end{tabularx}
    \end{doublerule}
    \caption{Scaling dimensions and OPE ratios for the lowest dimensional scalars for $\Delta_1=0.755$ (corresponding to the $CP^2$ model) at various $\Lambda$ values, together with the \( \Lambda \to \infty \) fit. The values of $\Delta_0$ are rigorous lower bounds. The values of \( \Delta_{2} \) and the OPE ratios correspond to the point on the boundary of the 5-dimensional allowed region \( (\Delta_{0}, \Delta_{2}, \lambda_{000}/\lambda_{110}, \lambda_{220}/\lambda_{110}, \lambda_{112}/\lambda_{110}) \) obtained by minimizing \( \Delta_{0} \) at fixed \( \Delta_{1} \). The values of \( \Delta_{3} \) and \( \Delta_{4} \) are non‐rigorous estimates from truncated crossing equations.\label{tab-raw-n3}}
\end{table}

\FloatBarrier  

\section{Results for the $CP^{3}$ and $CP^{4}$ models.}\label{app:N4-5}

In this appendix we describe the results obtained for $CP^{3}$ and $CP^{4}$ models by analogous bootstrap methods, i.e.\ by setting $\Delta_1$ to $0.88$ and $1.005$ (as predicted by the large $N$ prediction).
In Tables~\ref{table-scalar-monopoles-n4} and~\ref{table-scalar-monopoles-n5} we show the prediction of the lowest operator dimensions for scalars with charge $q$ in the $CP^{3}$ and $CP^{4}$ models, respectively, and compare these against large $N$ and lattice estimates.
The bootstrap estimate is obtained by extrapolating our finite $\Lambda$ results to infinite $\Lambda$ according to the linear fits shown in Figure~\ref{fig:fitplots-n4} and~\ref{fig:fitplots-n5}, as described in Appendix~\ref{app:N3}.
The data at finite values of $\Lambda$ is included in Tables~\ref{tab-raw-n4} and~\ref{tab-raw-n5} for \( CP^{3} \) and \( CP^{4} \), respectively.

\begin{table}[H]
    \renewcommand{\arraystretch}{1.2}
    \setlength{\tabcolsep}{5pt}
    \centering
    \begin{doublerule}
        \begin{tabularx}{\textwidth}{Y|@{\hskip 10 pt} XXXXX}
          $CP^3$ & $\Delta_1$ & $\Delta_2$ & $\Delta_{3}$ & $\Delta_{4}$ & $\Delta_{0}$ \\ \hline
          Bootstrap & \( 0.880^{*} \) & \( 2.12 (1)\) & \( 3.65 (4) \) & \( 5.8 (2) \) & \( 1.43 (2) \) \\
          Large $N$& \( 0.880 \) & \( 2.12 \) & \( 3.64 \) & \( 5.40 \)  & -- \\
          Lattice~\cite{2009PhRvB..80r0414L} &  \( 0.85(1) \) & -- & --  & -- &  \( 1.57(4) \)\\
          Lattice~\cite{2013PhRvB..88v0408H} &  \( 0.865 \) & -- & -- & --  & \( 1.60 \)
        \end{tabularx}
    \end{doublerule}
    \caption{Comparison of scaling dimensions $\Delta_q$ of the lowest dimension scalar operators with $U(1)$ charge $q$ in the \( CP^{3} \) model, as determined from the bootstrap study here, the large $N$ expansion for monopoles $q>0$, and lattice studies. The asterisk by $\Delta_1$ for bootstrap means we put it in to determine the others, while the bootstrap errors come from extrapolation in the bootstrap truncation parameter $\Lambda$.\label{table-scalar-monopoles-n4}}
\end{table}

\begin{table}[H]
    \centering
    \renewcommand{\arraystretch}{1.2}
    \setlength{\tabcolsep}{5pt}
    \begin{doublerule}
        \begin{tabularx}{\textwidth}{Y|@{\hskip 10 pt} XXXXXXX}
          $CP^3$
          & $\Lambda=19$ & $\Lambda=21$ & $\Lambda=23$ & $\Lambda=25$
          & $\Lambda=27$ & $\Lambda=29$ & $\Lambda \to \infty$ \\
          \hline
          $\Delta_0$
          & 1.303 & 1.317 & 1.329 & 1.336
          & 1.342 & 1.347 & 1.432 \\
          $\Delta_2$
          & 2.161 & 2.157 & 2.151 & 2.151
          & 2.148 & 2.149 & 2.123 \\
          $\Delta_3$
          & 3.766 & 3.753 & 3.737 & 3.735
          & 3.727 & 3.728 & 3.647 \\
          $\Delta_4$
          & 6.734 & 6.669 & 6.542 & 6.510
          & 6.436 & 6.435 & 5.795 \\
          $\lambda_{000}/\lambda_{110}$
          & 0.096 & 0.103 & 0.108 & 0.112
          & 0.115 & 0.118 & 0.161 \\
          $\lambda_{220}/\lambda_{110}$
          & 1.636 & 1.645 & 1.650 & 1.656
          & 1.658 & 1.662 & 1.711 \\
          $\lambda_{112}/\lambda_{110}$
          & 1.633 & 1.635 & 1.643 & 1.644
          & 1.647 & 1.648 & 1.681 \\
        \end{tabularx}
    \end{doublerule}
    \caption{Scaling dimensions and OPE ratios for the lowest dimensional scalars for $\Delta_1=0.88$ (corresponding to the $CP^{3}$ model) at various $\Lambda$ values, together with the \( \Lambda \to \infty \) fit. The values of $\Delta_0$ are rigorous lower bounds. The values of \( \Delta_{2} \) and the OPE ratios correspond to the point on the boundary of the 5-dimensional allowed region \( (\Delta_{0}, \Delta_{2}, \lambda_{000}/\lambda_{110}, \lambda_{220}/\lambda_{110}, \lambda_{112}/\lambda_{110}) \) obtained by minimizing \( \Delta_{0} \) at fixed \( \Delta_{1} \). The values of \( \Delta_{3} \) and \( \Delta_{4} \) are non‐rigorous estimates from truncated crossing equations.\label{tab-raw-n4}}
\end{table}

\begin{table}[H]
    \renewcommand{\arraystretch}{1.1}
    \setlength{\tabcolsep}{5pt}
    \centering
    \begin{doublerule}
        \begin{tabularx}{\textwidth}{c|@{\hskip 10 pt} XXXXX}
              $CP^4$& $\Delta_1$ & $\Delta_2$ & $\Delta_{3}$ & $\Delta_{4}$ & $\Delta_{0}$ \\ \hline
              Bootstrap & \( 1.005^{*} \) & \( 2.58 (2) \) & \( 4.6 (4) \) & \( 3.5(4) \) & \( 1.154(5) \) \\
              Large $N$& \( 1.005 \) & \( 2.43 \) & \( 4.18 \) & \( 6.21 \)  & -- \\
              Lattice rect.~\cite{2013PhRvL.111m7202B} & \( 0.85(10) \) & -- & --  & -- & \( 1.15 \) \\
              Lattice honeyc.~\cite{2013PhRvL.111m7202B} & \( 1.0(1) \) & -- & -- & --  & \( 1.46 \)
        \end{tabularx}
    \end{doublerule}
    \caption{Comparison of scaling dimensions $\Delta_q$ of the lowest dimension scalar operators with $U(1)$ charge $q$ in the \( CP^{4} \) model, as determined from the bootstrap study here, the large $N$ expansion for monopoles $q>0$, and lattice studies (on rectangular and honeycomb lattices). The asterisk by $\Delta_1$ for bootstrap means we put it in to determine the others, while the bootstrap errors come from extrapolation in the bootstrap truncation parameter $\Lambda$.\label{table-scalar-monopoles-n5}
	}
\end{table}

\begin{table}[hp]
    \centering
    \renewcommand{\arraystretch}{1.1}
    \setlength{\tabcolsep}{5pt}
    \begin{doublerule}
        \begin{tabularx}{\textwidth}{Y| @{\hskip 10 pt} XXXXXXXX}
          $CP^4$
          & $\Lambda=19$ & $\Lambda=21$ & $\Lambda=23$ & $\Lambda=25$
          & $\Lambda=27$ & $\Lambda=29$ & $\Lambda=31$ & $\Lambda \to \infty$ \\
          \hline
          $\Delta_0$
          & 1.063 & 1.071 & 1.080 & 1.086
          & 1.091 & 1.094 & 1.098 & 1.154 \\
          $\Delta_2$
          & 2.534 & 2.552 & 2.539 & 2.547
          & 2.545 & 2.554 & 2.553 & 2.576 \\
          $\Delta_3$
          & --    & 3.331 & --    & 4.575
          & 4.558 & 4.588 & 4.569 & 4.589 \\
          $\Delta_4$
          & --    & 4.129 & 10.348 & 3.787
          & 3.746 & 3.758 & 3.731 & 3.534 \\
          $\lambda_{000}/\lambda_{110}$
          & \( -0.830 \) & \( -0.918 \) & \( -0.972 \) & \( -1.051 \)
          & \( -1.072 \) & \( -1.131 \) & \( -1.106 \) & \( -1.615 \) \\
          $\lambda_{220}/\lambda_{110}$
          & 1.567 & 1.561 & 1.560 & 1.557
          & 1.563 & 1.568 & 1.569 & 1.571 \\
          $\lambda_{112}/\lambda_{110}$
          & 2.892 & 2.979 & 3.003 & 3.065
          & 3.058 & 3.099 & 3.065 & 3.396 \\
        \end{tabularx}
    \end{doublerule}
    \caption{Scaling dimensions and OPE ratios for the lowest dimensional scalars for $\Delta_1=0.1005$ (corresponding to the $CP^{4}$ model) at various $\Lambda$ values, together with the \( \Lambda \to \infty \) fit. The values of $\Delta_0$ are rigorous lower bounds. The values of \( \Delta_{2} \) and the OPE ratios correspond to the point on the boundary of the 5-dimensional allowed region \( (\Delta_{0}, \Delta_{2}, \lambda_{000}/\lambda_{110}, \lambda_{220}/\lambda_{110}, \lambda_{112}/\lambda_{110}) \) obtained by minimizing \( \Delta_{0} \) at fixed \( \Delta_{1} \). The values of \( \Delta_{3} \) and \( \Delta_{4} \) are non‐rigorous estimates from truncated crossing equations.\label{tab-raw-n5}}
\end{table}

\begin{figure}[hp]
	\centering
	\begin{tabularx}{\textwidth}{YYY}
		\includegraphics[width=0.325\textwidth]{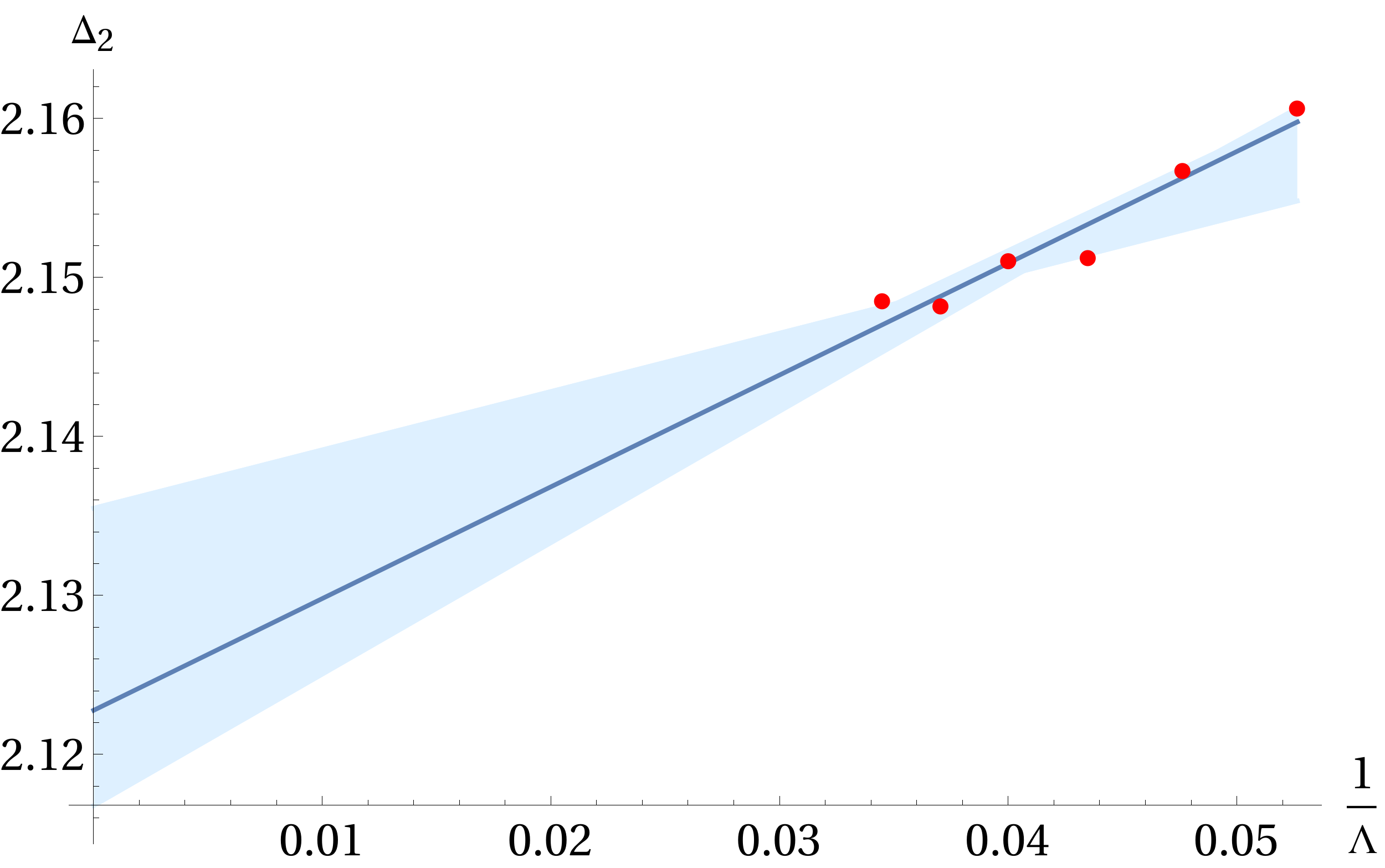} &
		\includegraphics[width=0.325\textwidth]{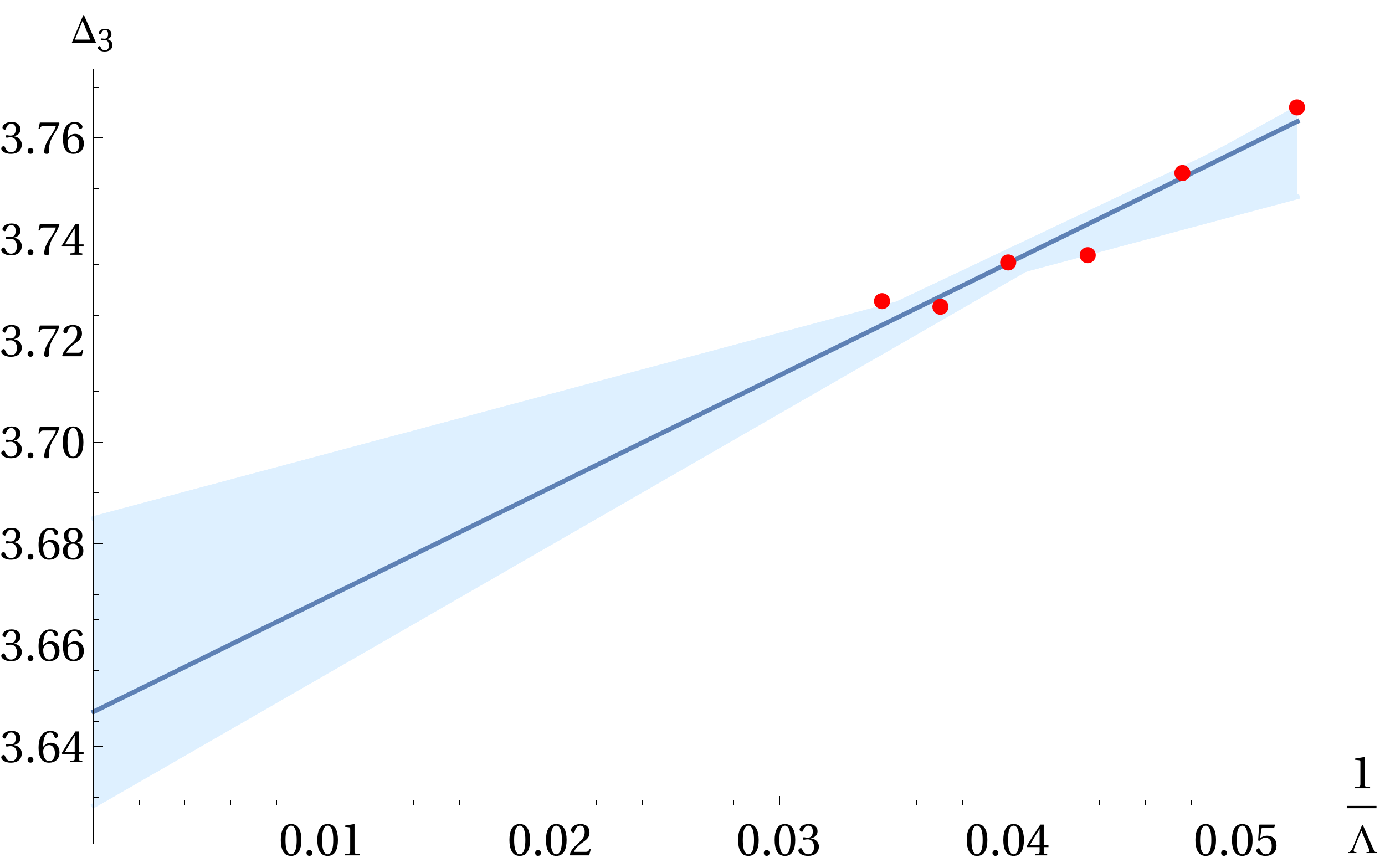} &
		\includegraphics[width=0.325\textwidth]{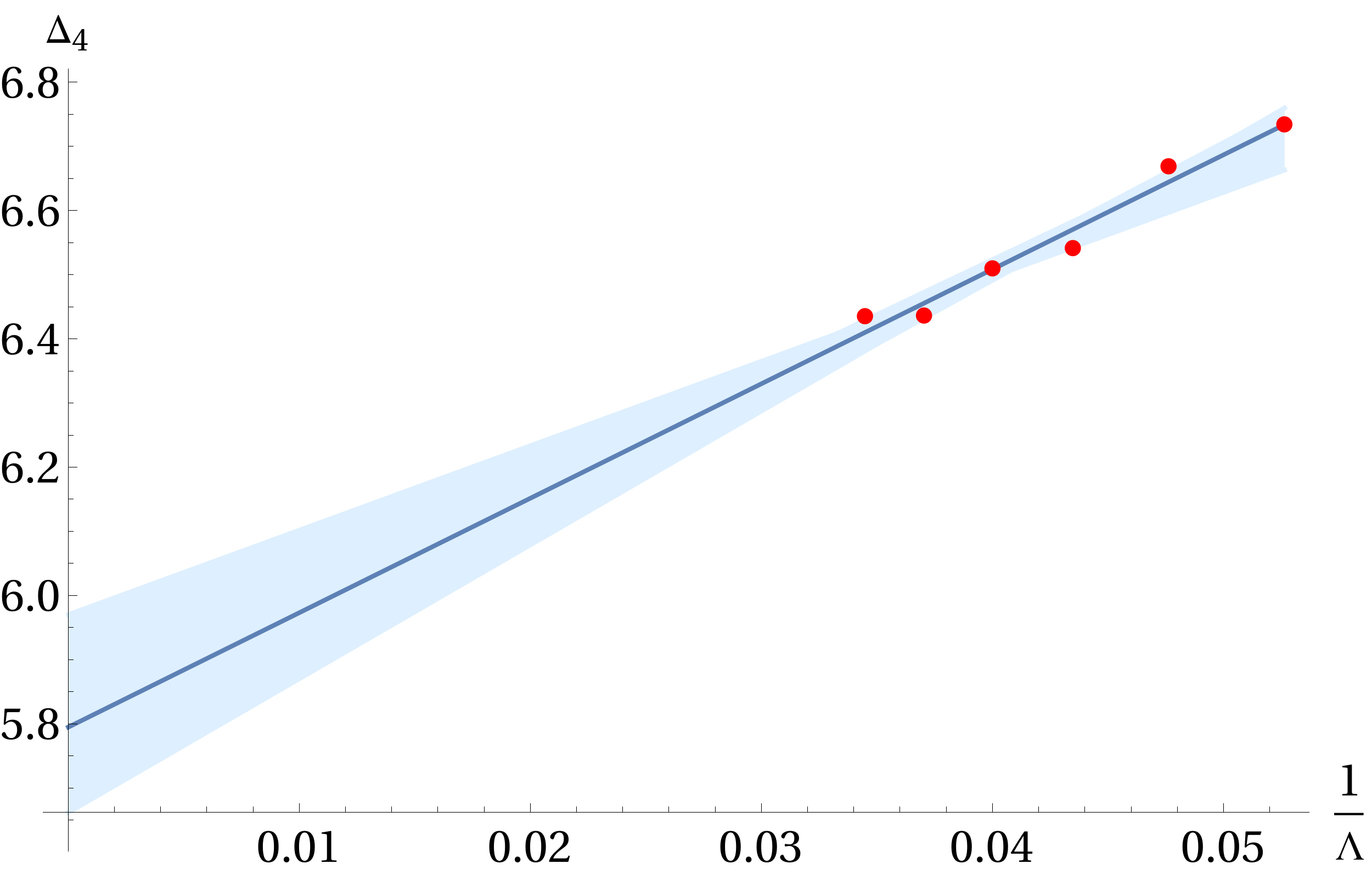}
	\end{tabularx}
	\caption{Linear least-squares fits in $1/\Lambda$ of the dimension of the lowest scalar monopoles in the $CP^{3}$ model (excluding ``fake'' operators at the imposed gap, when present). These fits are used to obtain the extrapolations at infinite $\Lambda$ shown in Table~\ref{table-scalar-monopoles-n4}. To give an estimate of the stability of the fit and its predictive accuracy a blue band indicate the range of fits that can be obtained by omitting up to 2 points. The red dots indicate the dimensions (obtained by EFM except for $\Delta_2$ which is one of the variables in the search space).\label{fig:fitplots-n4}}
\end{figure}

\begin{figure}[hp]
	\centering
	\begin{tabularx}{\textwidth}{YYY}
		\includegraphics[width=0.325\textwidth]{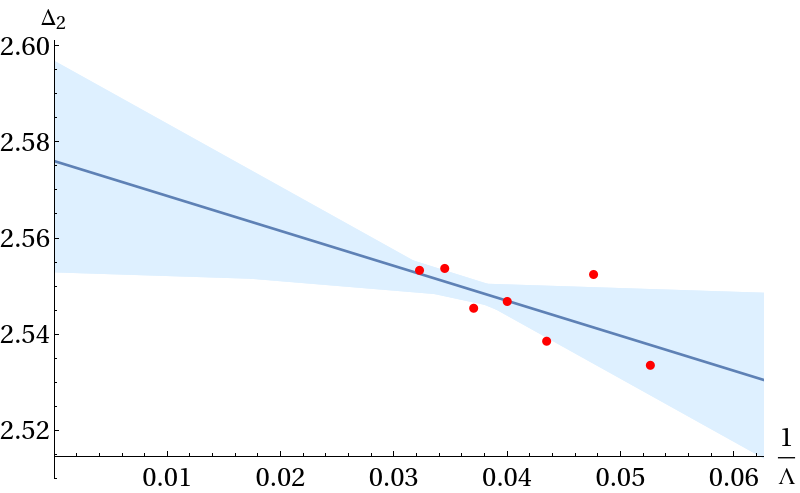} &
		\includegraphics[width=0.325\textwidth]{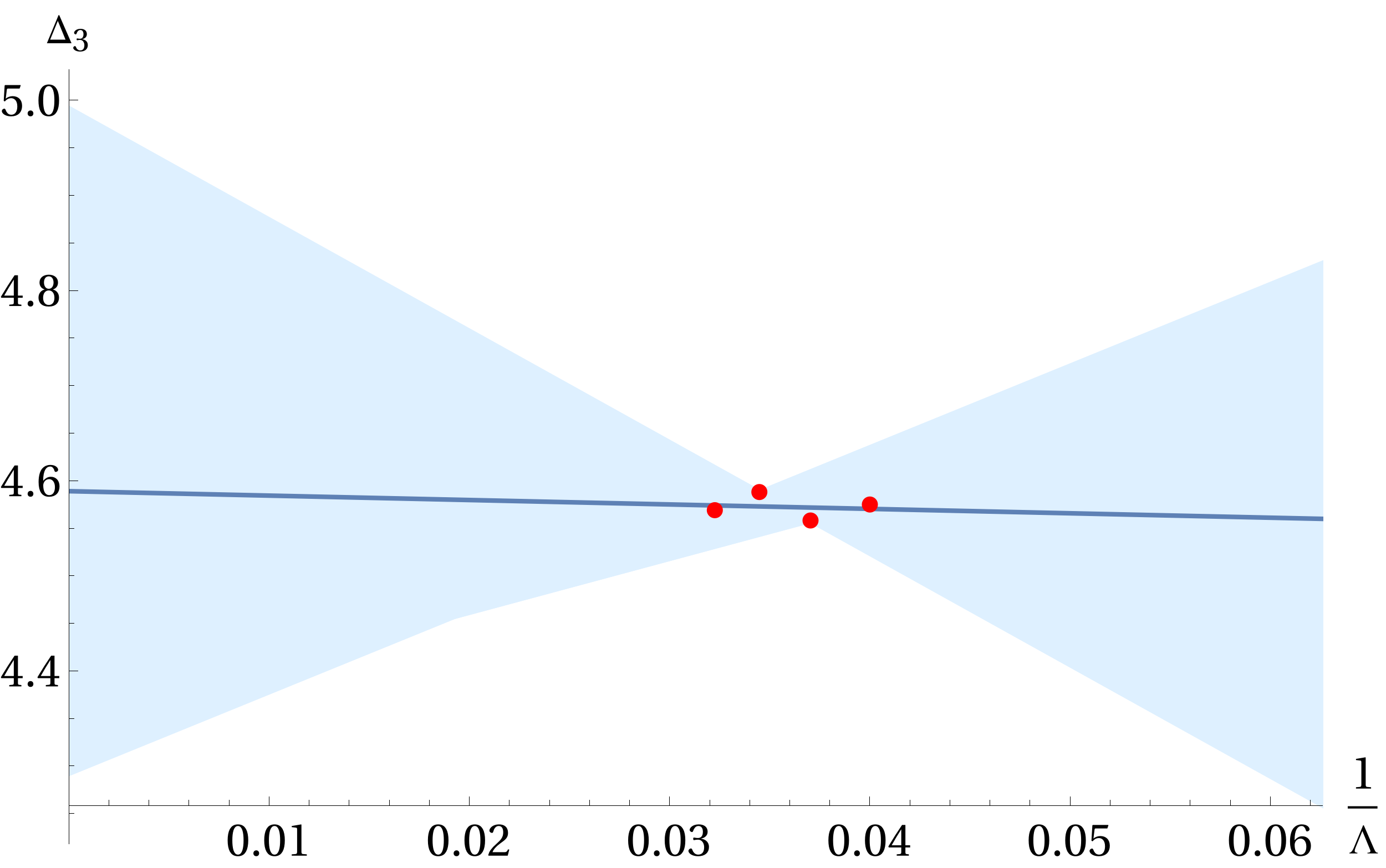} &
		\includegraphics[width=0.325\textwidth]{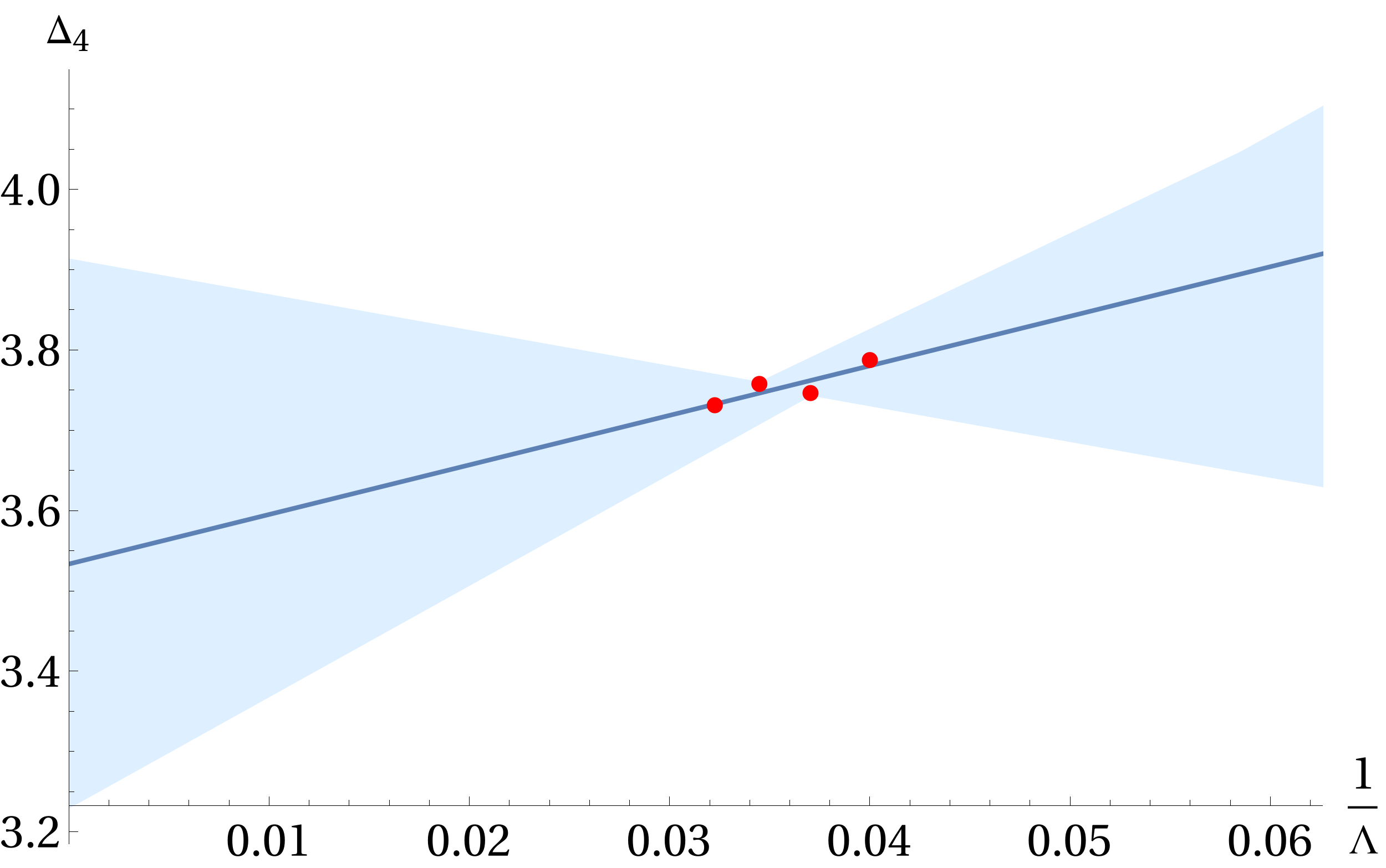}
	\end{tabularx}
	\caption{Linear least-squares fits in $1/\Lambda$ of the dimension of the lowest scalar monopoles in the $CP^{4}$ model (excluding ``fake'' operators at the imposed gap, when present). These fits are used to obtain the extrapolations at infinite $\Lambda$ shown in Table~\ref{table-scalar-monopoles-n5}. To give an estimate of the stability of the fit and its predictive accuracy a blue band indicate the range of fits that can be obtained by omitting up to 2 points. The red dots indicate the dimensions (obtained by EFM except for $\Delta_2$ which is one of the variables in the search space). Some figures contain less points since some operators are not detected at for all values of $\Lambda$.\label{fig:fitplots-n5}
    }
\end{figure}


\clearpage

\FloatBarrier

\section{Comparison of bootstrap to large $q$ for critical $O(2)$}
\label{app:o2}

In~\cite{Banerjee:2017fcx}, the scaling dimensions of scalar operators in the critical $O(2)$ model were computed for $q\leq16$ using lattice Monte Carlo. These values were fit to the general large $q$ expansion in~\eqref{largeq} for $\ell=0$ to get the coefficients
\es{c1c2app}{
c_{\frac32}=0.3371\,,\qquad c_{\frac12}=0.2658\,.
}
We show the predictions from~\eqref{largeq} with these value of the coefficients for $q=1,2,3,4$ and $\ell = 0, 2, 3, 4$ in Table~\ref{largeqtab}. Note that just as for the $CP^2$ model, for some values of $\ell>q$ we observe violations of the unitarity bound, which suggests the expansion is not good for these small values of $q$.

The conformal bootstrap was used in~\cite{Liu:2020tpf} to compute scaling dimensions of operators in the critical $O(2)$, using the same set of correlators of $q=0,1,2$ scalar operators considered in this paper, except with assumptions on the spectrum suitable to the critical $O(2)$ model. In Table~\ref{o2boot}, we summarize the data from this bootstrap study for $q=1,2,3,4$ and $\ell\leq4$, where recall that odd spins do not appear for $q=4$ in the correlators considered in this bootstrap. Some of this data is known to more digits of precision than shown here, but for the purpose of the comparison to large $q$, we find that the 4 digits is more than enough. As one can see from comparing Tables~\ref{largeqtab} and~\ref{o2boot}, the data from each method matches for $\ell\leq q$.

Note that \cite{Liu:2020tpf} did not give data for $q=3$ odd $\ell$ or $q=\ell=1$, so we reran those values ourselves.

\begin{table}[h]
    \centering
    \renewcommand{\arraystretch}{1.2}
    \setlength{\tabcolsep}{5pt}
    \begin{doublerule}
        \begin{tabularx}{\textwidth}{Y |  @{\hskip 10 pt} XXXX}
          Large charge  &$\ell=0$& $\ell=2$ & $\ell=3$ & $\ell=4$ \\ \hline
          $q=1$ & $0.509$  & ${\red2.241}$ & ${\red2.958}$ & ${\red3.672}$  \\
          $q=2$ & $1.236$  & $2.968$ & ${\red3.685}$ & ${\red4.399}$  \\
          $q=3$ & $2.118$  & $3.851$ & $4.568$ & ${\red5.281}$  \\
          $q=4$  & $3.135$  & $4.867$ & $5.584$& $6.297$ \\
        \end{tabularx}
    \end{doublerule}
    \caption{Scaling dimensions of the lowest dimension scalar operators in the critical $O(2)$ model as computed from the large $q$ effective theory for general $\ell\neq1$ with coefficients fixed by Monte Carlo for $\ell=0$.
        The red values for $\ell>q$ are where this small $\ell$ expansion is observed to be less accurate.\label{largeqtab}}
\end{table}

\begin{table}[h]
    \centering
    \renewcommand{\arraystretch}{1.2}
    \setlength{\tabcolsep}{5pt}
    \begin{doublerule}
        \begin{tabularx}{\textwidth}{Y|@{\hskip 10pt} XXXXX}
          Bootstrap & $\ell=0$ & $\ell=1$ & $\ell=2$ & $\ell=3$ & $\ell=4$ \\
          \hline
          $q=1$  & $0.519$ & $2.950$ & $3.650$ & $4.615$ & $5.700$ \\
          $q=2$  & $1.236$ & $5.800$ & $3.015$ & $5.766$ & $5.030$ \\
          $q=3$  & $2.106$ & $2.078$ & $3.884$ & $4.582$ & $5.852$ \\
          $q=4$  & $3.115$ & -- & $4.893$ & -- & $6.730$  \\
        \end{tabularx}
    \end{doublerule}
    \caption{Scaling dimensions of the lowest dimension scalar operators in the critical $O(2)$ model as computed from the conformal bootstrap.\label{o2boot}}
\end{table}

\end{document}